\documentclass[10pt]{article}
\usepackage{epsf,amsmath,amscd,amssymb,graphicx
}

\makeatletter
\@addtoreset{equation}{section}

\renewcommand{\thefootnote}{\fnsymbol{footnote}}
\makeatother

\newcommand\be{\begin{equation}}
\newcommand\ee{\end{equation}}
\newcommand\bea{\begin{eqnarray}}
\newcommand\eea{\end{eqnarray}}

\def\bd#1{\mbox{\boldmath $#1$}}
\begin{document}

\thispagestyle{empty}
\begin{flushright}
TIT/HEP--500 \\
{\tt hep-th/0306198} \\
June, 2003 \\
\end{flushright}
\vspace{3mm}
\begin{center}
{\Large
{\bf BPS Multi-Walls 
in Five-Dimensional Supergravity}
\\
} 
\vskip 1.5cm

  {\large \bf 
  Minoru~Eto}
\footnote{\it  e-mail address: 
meto@th.phys.titech.ac.jp
},  
 {\large \bf 
Shigeo~Fujita}
\footnote{\it  e-mail address: 
fujita@th.phys.titech.ac.jp
},
  {\large \bf 
  Masashi~Naganuma}
\footnote{\it  e-mail address: 
naganuma@th.phys.titech.ac.jp
},  

~and~~  {\large \bf 
Norisuke~Sakai}
\footnote{\it  e-mail address: 
nsakai@th.phys.titech.ac.jp
}

\vskip 1.5em

{ \it 
Department of Physics, Tokyo Institute of Technology 
\\
Tokyo 152-8551, JAPAN  }
\vspace{10mm}
{\bf Abstract}\\[5mm]
{\parbox{14cm}{\hspace{5mm}
Exact BPS solutions of multi-walls are obtained in 
five-dimensional supergravity. 
The solutions contain $2n$ parameters 
similarly to the moduli space 
of the corresponding global SUSY models and 
have a smooth limit of vanishing gravitational 
coupling. 
The models are constructed as gravitational deformations of 
massive $T^*{\bf C}P^n$ nonlinear sigma models 
by using the off-shell formulation of supergravity 
and massive quaternionic quotient method with 
$U(1) \times U(1)$ gauging. 
We show that the warp factor can have at most single 
stationary point in this case. 
We also obtain BPS multi-wall solutions even for 
models which reduce 
to generalizations 
of massive $T^*{\bf C}P^n$ 
models with only $U(1)^n$ isometry 
in the limit of vanishing gravitational 
coupling. 
At particular values of parameters, 
isometry of the quaternionic manifolds is enhanced. 
}}
\end{center}
\vfill
\newpage
\setcounter{page}{1}
\setcounter{footnote}{0}
\renewcommand{\thefootnote}{\arabic{footnote}}

\section{Introduction}\label{INTRO}

In the  brane world scenario, 
 our four-dimensional spacetime 
is realized on topological defects such as walls 
 embedded in a higher dimensional spacetime \cite{LED}. 
One of the most intriguing models 
in the brane-world scenario is proposed 
by Randall and Sundrum, which offers a possibility 
to solve the gauge hierarchy problem 
 by means of two walls \cite{RS1}, 
 and a four-dimensional localized massless graviton 
on the wall \cite{RS2}. 
The model is based on a spacetime metric 
containing a warp factor ${\rm e}^{2U(y)}$ 
\begin{eqnarray}
  ds^2=g_{\mu \nu}dx^\mu dx^\nu 
  = {\rm e}^{2U(y)}\eta_{mn}dx^mdx^n + dy^2,
\label{5dmetric}
\end{eqnarray}
where the five-dimensional indices transforming 
under the general 
coordinate transformations are denoted by 
$\mu, \nu = 0,1,2,3,4$, the flat four-dimensional 
indices $m, n = 0,1,3,4$. 
We choose to denote the coordinate of the extra dimension 
$y\equiv x^2$. 
They had to introduce an orbifold, a bulk cosmological 
constant and a boundary 
cosmological constant placed at orbifold 
fixed points. 
A fine-tuning was needed between these cosmological 
constants.

On the other hand, supersymmetry (SUSY) has been 
most powerful to construct realistic models 
for unified theories beyond the standard model \cite{DGSW}. 
The supersymmetric version of the warped metric model 
with an orbifold 
has been worked out \cite{ABN}. 
It is tempting to replace the orbifold by a smooth 
wall constructed from physical scalar fields. 
After many studies of BPS walls in four-dimensional 
${\cal N}=1$ supergravity (SUGRA) 
coupled with  chiral scalar 
multiplets \cite{CQR}, an 
exact solution of a BPS wall has been constructed 
\cite{EMSS}. 
Since the nonlinear models in the ${\cal N}=2$ 
SUGRA should have a quarternioinic K\"ahler 
target manifold  \cite{BaggerWitten}, there must 
be a nontrivial 
gravitational deformation of the hyper-K\"ahler nonlinear 
sigma models of the global ${\cal N}=2$ SUSY theories. 
 For massless ${\cal N}=2$
nonlinear sigma models in four dimensions, 
the necessary gravitational deformations 
have been worked out \cite{Galicki}, 
\cite{IvanovValent}. 
However, nontrivial potential terms are needed 
to obtain domain wall solutions. 
In the ${\cal N}=2$ SUSY nonlinear sigma models, 
the possible potential terms are severely restricted 
\cite{GTT1}--\cite{To2}. 
Massive hyper-K\"ahler nonlinear sigma models without 
gravity in four dimensions have been constructed 
in harmonic superspace as well as in ${\cal N}=1$ superfield 
formulation \cite{ANNS}, \cite{ANNS2}, 
and have yielded the domain wall 
solution for the 
Eguchi-Hanson manifold \cite{Eguchi} previously obtained 
in the on-shell component formulation \cite{GPTT}. 
Other BPS solitons in ${\cal N}=2$ global 
SUSY theories such as 
lumps have also been 
constructed \cite{AbrahamTownsend2}, \cite{NNS1}. 

Studies of domain walls in gauged SUGRA theories 
in five dimensions \cite{SUGRAwall} 
revealed the necessity of hypermultiplets 
\cite{KalloshLinde}. 
Domain walls in massive quaternionic K\"ahler nonlinear 
sigma models 
in the ${\cal N}=2$ SUGRA theories have been 
studied using 
mostly homogeneous target manifolds. 
Unfortunately, SUSY vacua in homogeneous target manifolds 
are not truly IR critical points, 
but can only be saddle points with some IR directions 
\cite{Alekseevsky}, \cite{BC2}. 
No-go theorems are argued \cite{MaldacenaNunez}, 
\cite{Gibbons-Lam}, 
 and solutions proposed \cite{Cvetic-Lam}. 
Only 
a few nonlinear sigma models with quaternionic manifolds 
as target spaces admitting wall solutions 
have been constructed 
\cite{Beh-Dall}, \cite{Lazaroiu}. 
However, a limit of weak 
gravitational coupling cannot be taken in these manifolds, 
contrary to the model with an 
exact solution in ${\cal N}=1$ SUGRA in 
four-dimensions \cite{EMSS}. 
This is a serious drawback for phenomenological purposes. 
Recently we have 
succeeded in constructing a smooth wall configuration 
which has a smooth limit of vanishing gravitational 
coupling \cite{AFNS}, using the recently developed off-shell 
formalism 
of ${\cal N}=2$ SUGRA in five dimensions 
\cite{Fujita-Ohashi}, \cite{FKO}. 
On-shell SUGRA in five dimensions 
\cite{Cere-Dall} 
as well as off-shell form without vector multiplets 
\cite{Zucker} 
have also been formulated previously. 

The purpose of our paper is to extend our construction of 
exact BPS domain walls in five-dimensional SUGRA
coupled with hypermultiplets (and vector multiplets), 
especially to higher dimensional quaternionic target spaces. 
We obtain a gravitational deformations of 
$T^*{\bf C}P^n$ 
nonlinear sigma models which admit multi-walls as BPS 
solutions. 
We shall call them $QT^*{\bf C}P^n$ models. 
These multi-wall solutions possess $2n$ parameters which is 
the same number as that of moduli parameters in the global 
SUSY case (vanishing gravitational coupling). 
We can take a limit of vanishing gravitational coupling 
smoothly as in our previous example of 
gravitational deformations of the 
$T^*{\bf C}P^1$ model (Eguchi-Hanson 
manifold) \cite{AFNS}. 
We also construct gravitational 
deformations with the $U(1) \times U(1)$ gauging 
of the hypermultiplets in the ${\cal N}=2$ SUGRA, 
which reduce to  more general models than the $T^*{\bf C}P^n$ 
model with only $U(1)^n$ isometry. 
We shall call them generalized $QT^*{\bf C}P^n$ models. 
We find that they admit multi-wall BPS solutions 
as well. 
It is also observed that the isometry of the target 
space is enhanced at particular values of parameters. 
Qualitative features of the wall solutions for 
hypermultiplets 
are found to be in one-to-one correspondence with 
those in the global SUSY case, even though there are 
nontrivial gravitational corrections 
\cite{GTT}. 
In addition, we obtain solutions of the warp factor. 
We find that possible stationary points of the warp 
factor are obtained as intersections with a 
hypersurface in field space and are easily 
visualized graphically. 
We show that the warp factor in the gravitationary dressed 
$T^*{\bf C}P^n$ models can have at most one stationary 
point. 

Similarly to our previous example, we obtain three types of 
behavior of the warp factor : 
1) decreasing 
for both infinities of extra dimension 
$y\rightarrow \pm\infty$, 
interpolating two infrared (IR) fixed points, 
2) decreasing in one direction, and flat 
 in the other, interpolating an IR fixed point 
 and flat space, 
3) decreasing in one direction, and 
 increasing in the other, interpolating an IR and an 
 ultra-violet (UV) fixed points. 
The IR-IR case may be desirable for phenomenological 
purposes \cite{Behrndt}, 
but other cases are also interesting in view of the AdS/CFT 
correspondence \cite{AdS/CFT}. 

We can take a thin-wall limit of our solitons 
(domain walls) in five-dimensional SUGRA
which gives orbifold type models 
similar to the Randall-Sundrum model. 
We find that the relation between bulk and boundary 
cosmological constants is now 
realized as an automatic consequence of the solution 
of dynamical equations rather than a fine-tuning 
between input parameters \cite{AFNS}, 
similarly to the BPS wall solution in the four-dimensional 
SUGRA \cite{EMSS}. 
The four-dimensional graviton should be localized also 
on our wall solution \cite{CsabaCsaki}. 

In constructing 
a gravitational deformation of nonlinear sigma models, 
we followed the strategy of our previous paper 
by using the 
off-shell formulation of five-dimensional SUGRA
(tensor calculus) \cite{Fujita-Ohashi}, \cite{FKO}. 
We combine this formalism with the quotient method 
via a vector multiplet without kinetic term and the 
massive deformation (central charge extension). 
In the off-shell formulation, we can easily introduce 
the gravitational coupling to the massive hypermultiplets 
with linear kinetic term which is interacting with the 
vector multiplet without kinetic term. 
By eliminating the vector multiplet 
after coupling to gravity, 
we automatically obtain a gravitationally deformed constraint 
resulting in inhomogeneous quaternionic 
K\"ahler nonlinear sigma model with the necessary potential 
terms. 
We may call the procedure a {\it massive} quaternionic 
K\"ahler quotient method. 
As an extension to our previous model \cite{AFNS},
we introduce many more 
hypermultiplets, but retain the same number of 
vector multiplets to gauge the hypermultiplet 
target space. 
Therefore we obtain higher dimensional quaternionic 
K\"ahler manifolds as target space. 
Although we have obtained legitimate quaternionic manifolds 
by construction at least locally, we have not yet explored 
possible global singularities which was noticed in the 
case of the $T^*{\bf C}P^1$ model (Eguchi-Hanson 
manifold) \cite{IvanovValent}, \cite{AFNS}. 
As we have argued before \cite{AFNS}, we can restore the 
kinetic term of the vector multiplet to avoid such 
singularities if they exist. 
Our BPS solutions should still 
be valid solutions even with the kinetic term for the 
vector multiplet.

Sec.\ref{sc:action} summarizes the massive quaternionic 
K\"ahler quotient method 
in the off-shell formulation of 
SUGRA, and introduces the general $U(1)\times U(1)$ 
gauging leading to $4n$ (real) dimensional 
quaternionic nonlinear sigma models. 
Sec.\ref{sc:BPSequation} gives vacua and BPS equations 
of the model. 
Sec.\ref{sc:exactBPSsol} discusses general properties of 
BPS solutions and presents exact multi-wall solutions. 
Sec.\ref{sc:discuss} 
discusses the weak gravity limit and other issues. 

\section{Our model} 
\label{sc:action}

\subsection{  
Five-dimensional Supergravity
} 

Our model is based on the off-shell formulation of 
Poincar\'e SUGRA in five-dimensions with 
 hypermultiplets and vector multiplets. 
The Poincar\'e SUGRA
is most conveniently obtained from 
a gauge fixing of superconformal gravity 
 \cite{Fujita-Ohashi}, which comes from 
gauging the translation $P_a$, supersymmetry $Q_i$, 
local Lorentz transformation $M_{ab}$, dilatation $D$, 
$SU(2)_R$ 
transformation $U_{ij}$, conformal supersymmetry $S_i$ and 
special conformal boost $K_a$ where $a=0, \dots, 4$ 
denote local Lorentz indices in five dimensions, 
and $i, j=1,2$ denote $SU(2)_R$ indices. 
Some of these gauge 
fields are written in terms of other fields 
(become auxiliary fields) after imposing appropriate 
constraints, and the resulting superconformal gravity 
contains the Weyl multiplet consisting of 
vielbein $e_\mu {}^a$, gravitino $\psi_\mu^i$, 
$SU(2)_R$ gauge field $V_\mu^{ij}$, and the additional 
fields to balance the bosonic and fermionic degrees of 
freedom, namely a real antisymmetric tensor field 
$v_{ab}$, an $SU(2)$-Majorana spinor field 
$\chi^i$, and a real scalar field $D$. 
Altogether the Weyl multiplet consists of the $32$ 
bosonic and $32$ fermionic components 
\cite{Fujita-Ohashi}, \cite{FKO}\footnote{ 
We adopt the conventions of Ref.\cite{Fujita-Ohashi} except 
the sign of our metric being 
$\eta_{\mu\nu}= diag(-1,+1,+1,+1,+1)$. 
This induces a change of the convention of 
Dirac matrices 
and the form of SUSY transformation of fermion.
}.

In global ${\cal N}=2$ SUSY case, 
BPS wall solutions have been obtained 
by using $n+1$ \; ($n \ge 1$) hypermultiplets as matter 
fields \cite{GPTT}, \cite{GTT}, \cite{ANNS}, \cite{ANNS2}. 
To obtain nontrivial interactions among hypermultiplets, 
the nonlinearity of the kinetic term (SUSY nonlinear 
sigma model) is required \cite{AF2}. 
The most practical way to obtain the nonlinear sigma 
model is to introduce a vector multiplet 
without a kinetic term that provides constraints 
to hypermultiplets. 
After solving the constraints, one obtains the curved 
target space for hypermultiplets that becomes a 
hyper-K\"ahler manifold with real $4n$ dimension 
 \cite{RT}, \cite{LR}, \cite{ANNS}, \cite{ANNS2}. 
To obtain wall solutions, we need also a nontrivial 
potential term among hypermultiplets. 
This comes about if we consider the central charge 
extension in the SUSY algebra producing the mass term 
\cite{SierraTownsend}. 
The above method to obtain the nonlinear sigma model 
with a potential term is called the massive 
hyper-K\"ahler quotient method \cite{ANNS}, \cite{ANNS2}. 

To embed such a massive SUSY nonlinear sigma model 
into SUGRA, we need to introduce an additional 
hypermultiplet (with a negative metric) that is called 
the compensator, since the gauge 
fixing in conformal supergravity (dilatation, $SU(2)_R$ 
and so forth) eliminates degrees of freedom of one 
hypermultiplet. 
The nontrivial potential term is obtained by introducing 
the mass term through the central charge extension.
This central charge extension in SUGRA
should be performed through an introduction of central charge 
vector multiplet  \cite{Fujita-Ohashi}, 
\cite{FKO} with the $U(1)$ gauge boson 
$W_\mu^0$, the generator $t_0$ and the gauge coupling 
$g_0$. 
After the gauge fixing from conformal supergravity, 
the resulting target space of 
physical hypermultiplets in Poincar\'e SUGRA
becomes a projective quaternionic 
manifold in contrast to the flat target space in the 
case of global SUSY theories \cite{Fujita-Ohashi}. 
To obtain a curved target space of nonlinear sigma models 
in the globally SUSY case, 
we have introduced a vector multiplet 
without kinetic term to serve as a Lagrange multiplier 
field \cite{ANNS}, \cite{ANNS2}. 
Similarly, the constraint on 
hypermultiplets is introduced in SUGRA theories by a 
vector multiplet without kinetic term, whose gauge boson, 
generator 
and coupling are denoted by $W_\mu^1$, $t_1$ and $g_1$. 
In this paper we introduce $U(1)$ gauge field as $W_\mu^1$.
Consequently, we need to introduce 
$U(1)\times U(1)$ vector multiplets 
besides the Weyl multiplets and $n+2$ hypermultiplets. 

The $n+2$ hypermultiplets can be assembled into 
a $2$ by $2(n+2)$ matrix whose scalar components are 
denoted as ${\cal A}_i{^\alpha}, \; i=1, 2 $. 
The hypermultiplet target space indices are denoted 
as $\alpha=1, \dots, 2(n+2)$. 
We have $4(n+2)$ real degrees of freedom in 
scalar fields ${\cal A}_i{}^\alpha$, 
because of the reality condition \cite{Fujita-Ohashi} 
\begin{equation}
{\cal A}^i{}_\alpha \equiv  \epsilon^{ij}{\cal A}_j{}^\beta 
\rho_{\beta \alpha} = - ({\cal A}_i{}^\alpha)^* ,
\qquad 
\rho_{\alpha\beta}
=\rho^{\alpha\beta}
= \left(
\begin{array}{ccc}
i\sigma_2 &  & \\
 &  & {\bf{1}}_{n+1} \\
 & -{\bf{1}}_{n+1} &  
\end{array}
\right),  
\end{equation}
where $\epsilon^{ij}=i\sigma_2$ and ${\bf{1}}_{n+1}$ 
is the $(n+1) \times (n+1)$ unit matrix.

After integrating out a part of the auxiliary fields 
by their on-shell conditions in 
the off-shell SUGRA action \cite{FKO}, 
we obtain the bosonic part of the action 
for our model 
\begin{eqnarray}
   e^{-1}{\cal L}
   &\!\!\!=&\!\!\!
  -\frac{1}{2\kappa^2}R-\frac{1}{4}
F_{\mu\nu}^0 F^{ 0\mu\nu} 
+ {c \over 8} \epsilon^{\lambda\mu\nu\rho\sigma}
W_\lambda^0F_{\mu\nu}^0 F_{\rho\sigma}^{ 0} 
\nonumber \\
   &\!\!\!{}&\!\!\! 
-\nabla^a{\cal A}_i{^{\bar{\alpha}}}\nabla_a
{\cal A}^i{_\alpha}
    - \kappa^2[{\cal A}_i{^{\bar{\alpha}}}
    \nabla_a{\cal A}^j{_\alpha}]
    [{\cal A}^{i{\bar{\beta}}}
    \nabla^a{\cal A}_{j\beta}] 
       \nonumber \\
   &\!\!\!{}&\!\!\! +{\cal A}_i{^{\bar{\alpha}}}
           \left[(g_IM^It_I)^2\right]_\alpha{}^\beta
           {\cal A}^i{}_\beta
          +\frac{\kappa^2}{12}(M^I{\cal Y}_I{^{ij}})
          (M^I{\cal Y}_{I{ij}}),
\label{SUGRA1}
\end{eqnarray}
where
\begin{eqnarray}
\begin{array}{l}
{\cal A}_i{^{\bar{\alpha}}} 
\equiv {\cal A}_i{^\beta}d_\beta{^\alpha},\\
\nabla_\mu{\cal A}_i{^\alpha}
  = \partial_\mu {\cal A}_i{^\alpha}
      -(g_IW^I_\mu t_I)^\alpha{}_\beta
      {\cal A}_i{}^\beta,\\
F_{\mu\nu}^0   = 
  \partial_\mu W_\nu^0-\partial_\nu W_\mu^0 ,\\
{\cal Y}_I{^{ij}} = 2g_I{\cal A}^{(i}{_\alpha}
d^\alpha{_\beta}
(t_I)^{\beta\gamma}{\cal A}^{j)}{_\gamma}.
\end{array}
\end{eqnarray}
Here, we follow the notation that 
 $d_\alpha{}^\beta = 
 diag(\bd{1}_2,-\bd{1}_{n+1},-\bd{1}_{n+1})$, 
$\kappa$ is the five-dimensional gravitational coupling,
$c$ is the coefficient of the trilinear `norm function' 
which appears in the Chern-Simons term and $I$ runs $0,1$.
The $M_I$ denote scalar fields in vector multiplets and 
the auxiliary fields ${\cal Y}_I{^{ij}}$ in vector 
multiplets are determined by their equations of motion. 
We drop the kinetic term of $W^1_\mu$ here\footnote{
If we turn to a notation such that  the gauge coupling $g_1$ 
is absorbed into a normalization of 
$W_\mu^1$ and take $g_1 \rightarrow \infty$ with 
$W_\mu^1$ fixed, the kinetic term of $W^1_\mu$ is 
dropped out of Lagrangian.
}. 
Moreover, this gives a constraint on hypermultiplet 
target space through the on-shell condition of the 
auxiliary field ${\cal Y}_1{^{ij}}$:
\begin{eqnarray}
  {1 \over g_1}{\cal Y}_1{^{ij}}
  & = & 2{\cal A}^{(i}{_\alpha}d^\alpha{}_\gamma
   (t_1)^{\gamma \beta}{\cal A}^{j)}{_\beta}
 \rightarrow 0 \qquad
\left(g_1\rightarrow\infty\right).
\label{2ndconst}
\end{eqnarray}
This constraint (\ref{2ndconst}) corresponds to the 
constraint for Eguchi-Hanson target space 
in the limit of $\kappa\to 0$ in the $n=1$ case 
Ref.\cite{ANNS}, \cite{ANNS2}. 
The gauge fixing of dilatation in 
superconformal gravity gives an additional constraint on 
hypermultiplet scalars as (with ${\cal A}_i{^{\bar{\alpha}}} 
\equiv {\cal A}_i{^\beta}d_\beta{^\alpha}$ in Eq.(2.3))
\begin{eqnarray}
  {\cal A}^2&\equiv&{\cal A}_i{^{\bar\alpha}}
  {\cal A}^i{_\alpha}
   = -2 \kappa^{-2} . 
\label{1stconst} 
\end{eqnarray}

Because of the quadratic form constraint (\ref{1stconst}), 
the target space manifold becomes 
$USp(2,2(n+1))/(U(2)\times U(2(n+1)))$ 
that has an isometry 
$USp(2,2(n+1))$ 
which is defined by the invariance of the 
quadratic form (\ref{1stconst}) \cite{Fujita-Ohashi}, 
\begin{eqnarray}
t^\dagger d + dt = 0, \qquad 
t^{\rm T}\rho - \rho t^\dagger=0,
\label{eq:constr-t}
\end{eqnarray}
where $t$ denotes the generator of the isometry group.
The general form of
$t$ satisfying these constraints is given by : 
\begin{eqnarray}
t^\alpha{_\beta} = \left(
\begin{array}{ccc}
\bd{A} & \bd{B}^\dagger & - \varepsilon\bd{B}^{\rm T}\\
\bd{B} & \tilde{ \bd{D}} & \bd{C}\\
\bd{B}^*\varepsilon & \bd{C}^* & - \tilde{\bd{D}}^{\rm T}
\end{array}
\right), 
\label{eq:USpgenerator}
\end{eqnarray}
where $\bd{A}$ is a $2$ by $2$ matrix satisfying, 
$\bd{A}^\dagger =-\bd{A}$ and 
$\bd{A}=\varepsilon\bd{A}^{\rm T}\varepsilon$, 
$\bd{B}$ is a $(n+1)$ by $2$ matrix, $\bd{C}$ is a 
$(n+1)$ by $(n+1)$ matrix satisfying 
$\bd{C}^{\rm T} = -\bd{C}$, and $\tilde{ \bd{D}}$ is 
a $(n+1)$ by $(n+1)$ matrix satisfying 
$\tilde{ \bd{D}}^\dagger = -\tilde{ \bd{D}}$. 

We should choose gauge generators $t_I$
for vector multiplets $W_\mu^I$ as one of the generators $t$
of the isometry $USp(2,2(n+1))$.
Since we are interested in $U(1)\times U(1)$ gauging, 
we choose two commuting diagonal generators parametrized by 
\begin{eqnarray}
\left(t_I\right)^\alpha{_\beta}
= \left(
\begin{array}{ccc}
i\alpha_I\sigma_3 &&\\
&i\bd{D}_I &\\
&&-i\bd{D}_I
\end{array}
\right),
\quad\quad\left(I=0,1\right)\label{t_I}
\end{eqnarray}
where $\sigma_3$ is one of the Pauli matrices, 
 $\bd{D}_I = diag(d_{I,1},d_{I,2},\cdots,d_{I,n+1})$, 
with $d_{I,a}$ and $\alpha_I$ are real parameters. 
We again stress that gauging by $t_1$ gives the 
non-minimal kinetic term for hypermultiplet scalar fields 
 ${\cal A}_i{}^\alpha$ through the 
 constraint. 
 Therefore the target manifold of the hypermultiplet 
 nonlinear 
 sigma model is entirely determined by the choice of $t_1$. 
On the other hand, the gauging by $t_0$ produces 
a nontrivial potential for hypermultiplet scalar 
fields ${\cal A}_i{}^\alpha$. 

Let us clarify the meaning of parameters $\alpha_I$ 
and $d_{I, \alpha}$. 
We wish to demand that our model should have a well-defined 
limit of vanishing gravitational coupling. 
The compensator hypermultiplet ${\cal A}_i{}^\alpha$ with 
$\alpha=1,2$ has negative norm and should disappear 
in the vanishing gravitational coupling. 
This requires $\alpha_I$ in $t_I$ should vanish 
in the limit  $\kappa\rightarrow0$. 
On the other hand, we wish to recover a 
nonlinear sigma model with the constraint given by 
the part of $(t_1)^\alpha{}_\beta$ with 
$\alpha, \beta=3, \cdots, 2(n+2)$. 
Similarly the potential for the hypermultiplets 
except the compensator is given by the part of 
$t_0$ with 
$\alpha, \beta=3, \cdots, 2(n+2)$. 
Therefore $d_{I,a}$ should have a finite limit 
as $\kappa\rightarrow0$. 
Even accepting these requirements, we still have a 
considerable freedom to choose parameters. 
For instance, 
a general choice for $\alpha_1$ and
$d_{1,a}$ makes 
the target manifold 
with only $U(1)^n$ isometry and in general 
inhomogeneous through the constraint (\ref{2ndconst}). 
We shall call this case as the {\it asymmetric kinetic term}.
We will consider BPS multi-wall solutions in 
such generic models. 
If we take a symmetric choice of parameters 
\begin{equation}
d_{1,a} =d_{1,n+1}, \qquad a=1,2,\cdots,n, 
\label{eq:sym-kin-term}
\end{equation}
the kinetic term becomes symmetric and 
the target manifold admits much larger 
isometry,  $SU(n+1)$ in this case. 
This is the case of the usual $T^*{\bf C}P^n$ 
nonlinear sigma model in the limit of vanishing 
gravitational coupling. 
We shall call this case as the {\it symmetric kinetic term}. 
On the other hand, we can obtain other target manifolds 
with enhanced isometry, 
if we abandon the weak gravity limit. 
We find that too much enhanced isometry often results 
in pathological behaviors such as runaway vacuum. 
A systematic search of possibilities for $n=1$ case 
is given in Appendix A. 

By reversing the sign of the third line of (\ref{SUGRA1}), 
we obtain a scalar potential consisting of two terms : 
the first term arises from the couplings of hypermultiplets 
to scalars $M^0, M^1$ in vector multiplets 
and the second term from eliminating the auxiliary fields 
of the $U(1)$ vector multiplet with kinetic term ($W_\mu^0$). 
The requirement of minimal kinetic term for the gauge field 
$W_\mu^0$ together with the canonical normalization 
of the Einstein-Hilbert term in Eq.(\ref{SUGRA1}) 
determine the value of 
the associated scalar $M^0$ as 
\begin{equation}
(M^0)^2=\frac{3}{2}\kappa^{-2}. 
\end{equation}
As we will see in later sections, the parameter $g_0M^0$ 
gives the scale of width of the wall and should be kept 
finite when taking the weak gravity limit 
$\kappa\rightarrow 0$. 
Therefore the gauge coupling $g_0$ of the central charge 
vector multiplet must be proportional to 
$\kappa$ for vanishing gravitational coupling. 
This fact implies that the central charge vector multiplet 
decouples in the global SUSY model. 
This result is in accord with the fact that 
the global SUSY model does not need the central 
charge vector multiplet to generate 
mass terms.

The scalar $M^1$ and the gauge field 
$W^1_\mu$ without kinetic term is a 
Lagrange multiplier, 
which is determined by its algebraic equations 
of motion as 
\begin{eqnarray}
 g_1M^1 &=& -\frac{{\cal A}_i{^{\bar\alpha}}  
     (t_0t_1)_\alpha{}^\beta{\cal A}^i{}_\beta}
     {{\cal A}_i{^{\bar\alpha}}[(t_1)^2]_\alpha{}^\beta
     {\cal A}^i{}_\beta}
g_0M^0 \equiv f({\cal A})g_0M^0,\label{M^1}\\
g_1W^1_\mu &=& \frac{1}{2}\frac{{\cal A}_i{^{\bar{\alpha}}}
\overset{\leftrightarrow}{\partial}_\mu t_{1\alpha}{}^\beta
{\cal A}^i{}_\beta
}{{\cal A}_i{^{\bar{\alpha}}}[(t_1)^2]_\alpha{}^\beta
{\cal A}^i{}_\beta}.\label{W^1}
\end{eqnarray}

\subsection{Bosonic action for Hypermultiplets}
Assuming $W_\mu^0=0$, 
we rewrite the on-shell bosonic action of the 
hypermultiplets:
\begin{eqnarray}
e^{-1}{\cal L}_{matter} 
&=& - \partial^\mu{\cal A}_i{^{\bar{\alpha}}}
\partial_\mu{\cal A}^i{_\alpha}
- \kappa^2\left({\cal A}^{i\bar{\alpha}}
\overset{\leftrightarrow}{\partial}_\mu
{\cal A}^j{_\alpha}\right)^2
+ (g_1W_\mu^1)^2 {\cal A}_i{^{\bar{\alpha}}}
\left(t_1{^2}\right)_\alpha{}^\beta{\cal A}^i_\beta
- V({\cal A}_i{^\alpha}),\\
V({\cal A}_i{^\alpha}) &=& 
-{\cal A}_i{^{\bar{\alpha}}}
\left[\left(g_IM^It_I\right)^2\right]_\alpha{}^\beta
{\cal A}^i{}_\beta
- \frac{\kappa^2}{12}
(M^0)^2\mathcal{Y}_0{^{ij}}\mathcal{Y}_{0{ij}},
\end{eqnarray}
where $M^1$ and $W^1_\mu$ are given in Eq.(\ref{M^1}) 
and (\ref{W^1}).
By using the $SU(2)_R$ gauge fixing, we can choose the first 
two columns in the hypermultiplet matrix 
${\cal A}_i{^\alpha}$ to be proportional to 
the two by two unit matrix. 
Remaining components of ${\cal A}_i{}^\alpha$ 
can be parametrized by 
two $(n+1)$-component complex fields 
$\phi_i{^a}$ with $i=1,2$ and $a=1,2,\cdots,n+1$
\begin{eqnarray}
 {\cal A}_i{}^\alpha \equiv \frac{1}{\kappa}
 \bar{\cal A}^{-1/2}\left(
\begin{array}{cccc}
1 & 0 & \kappa\bd{\phi}_1^T & -\kappa\bd{\phi}_2^{*T} \\
0 & 1 & \kappa\bd{\phi}_2^T & \kappa\bd{\phi}_1^{*T} \\
\end{array} \right),\quad
\bd{\phi}_i = \left(
\begin{array}{c}
\phi_i{^1}\\
\phi_i{^2}\\
\vdots\\
\phi_i{^{n+1}}
\end{array}
\right) = (\phi_i{^a}) .
 \label{CFbasis} 
\end{eqnarray}
\begin{equation}
\bar{\cal A}=1-\kappa^2(|\bd{\phi}_1|^2+|\bd{\phi}_2|^2)
\label{eq:dilatation-norm}
\end{equation}
which satisfies the constraint (\ref{1stconst}). 
This parametrization is a generalization of that used by 
Curtright and Freedman for $T^*{\bf C}P^1$ nonlinear sigma 
model ($n=1$ case) \cite{CF}. 
We shall call it the (generalized) Curtright-Freedman 
basis.

In terms of the Curtright-Freedman basis with the 
generators (\ref{t_I}),
$\mathcal{Y}_I{^{ij}}$ can be expressed as
\begin{eqnarray}
\mathcal{Y}_I{^{ij}} =
- 2g_I\bar{\cal A}^{-1}
\left(
\begin{array}{cc}
2i{\cal S}_I & i{\cal T}_I\\
i{\cal T}_I & -2i{\cal S}_I^*
\end{array}
\right).
\label{eq:Ymatrix}
\end{eqnarray}
where we define
\begin{eqnarray}
{\cal T}_I &\equiv& \kappa^{-2}\alpha_I -
\left(\bd{\phi}_1^\dagger\bd{D}_I\bd{\phi}_1 
- \bd{\phi}_2^\dagger\bd{D}_I\bd{\phi}_2\right),\\
{\cal S}_I &\equiv& 
\bd{\phi}_1^\dagger\bd{D}_I\bd{\phi}_2.
\end{eqnarray}
Hence, the constraint (\ref{2ndconst}) reduces to
\begin{eqnarray}
{\cal T}_1=0,\qquad {\cal S}_1 = {\cal S}_1^* = 0.
\label{2ndconst_2}
\end{eqnarray}

Finally, we obtain the bosonic Lagrangian 
in terms of the Curtright-Freedman basis as follows:
\begin{eqnarray}
\frac{1}{2}e^{-1}{\cal L}_{matter}
    &\!\!\!=&\!\!\! 
    -\bar{\cal A}^{-1}[|\partial_\mu \bd{\phi}_1|^2
            + |\partial_\mu \bd{\phi}_2|^2
  + (\kappa^{-2}\alpha_1^2 
  - \bd{\phi}_1^\dagger\bd{D}_1^2 \bd{\phi}_1 
  - \bd{\phi}_2^\dagger\bd{D}_1^2\bd{\phi}_2 )
             W^1_\mu W^{1\mu}]
         \nonumber \\
    &\!\!\!{}&\!\!\!\!-\kappa^2\bar{\cal A}^{-2}
          [|\bd{\phi}_2^\dagger\partial_\mu \bd{\phi}_2
           + \bd{\phi}_1^{\rm T}\partial_\mu \bd{\phi}_1^*|^2
           + |\bd{\phi}_1^{\rm T}\partial_\mu \bd{\phi}_2^*
           - \bd{\phi}_2^\dagger\partial_\mu \bd{\phi}_1|^2]
   - \frac{1}{2}V(\phi_i{^a}) ,
     \label{CFkin}\\
\frac{1}{2}V(\phi_i{^a}) &\!\!\!=&\!\!\!\! 
(g_0M^0)^2\bar{\cal A}^{-1}\left[-\kappa^{-2}\alpha^2 
\!
+
\! \left( \bd{\phi}_1^\dagger\bd{D}^2\bd{\phi}_1 
+ \bd{\phi}_2^\dagger\bd{D}^2\bd{\phi}_2\right)
\!-\! \frac{\kappa^2}{3}\bar{\cal A}^{-1}
\left({\cal T}_0^2 + 4|{\cal S}_0|^2\right)\!\right],
\label{CFpot}
\end{eqnarray}
where we define quantities combining two vector multiplets 
using $f$ in Eq.(\ref{M^1}) 
\begin{eqnarray}
\alpha \equiv \alpha_0 + f\alpha_1, 
\qquad 
d_a \equiv d_{0,a}+fd_{1,a}, 
\qquad 
\bd{D} \equiv \bd{D}_0 +f\bd{D}_1=diag\{d_a\}. 
\end{eqnarray}
We obtain the following useful expression:
\begin{eqnarray}
g_IM^It_I = g_0M^0(t_0 + ft_1)
= g_0M^0\left(
\begin{array}{ccc}
i\alpha\sigma_3 &&\\
& i\bd{D} & \\
&& -i\bd{D}
\end{array}
\right).
\end{eqnarray}
We can express $W^1_\mu$ and $f$ as follows:
\begin{eqnarray}
g_1W^1_\mu &=&  \frac{i}{2}\frac{
\bd{\phi}_1^\dagger\bd{D}_1
\overset{\leftrightarrow}{\partial}_\mu\bd{\phi}_1
+ \bd{\phi}_2^\dagger\bd{D}_1
\overset{\leftrightarrow}{\partial}_\mu\bd{\phi}_2
}{
\kappa^{-2}\alpha_1^2 - 
\left(
\bd{\phi}_1^\dagger\bd{D}_1^2
\bd{\phi}_1 + \bd{\phi}^\dagger_2\bd{D}_1^2\bd{\phi}_2
\right)
},\\
f &=& - \frac{\kappa^{-2}\alpha_0\alpha_1 
- \left(\bd{\phi}_1^\dagger\bd{D}_0\bd{D}_1\bd{\phi}_1 
+ \bd{\phi}_2^\dagger\bd{D}_0\bd{D}_1\bd{\phi}_2\right)}{
\kappa^{-2}\alpha_1 ^2
- \left(\bd{\phi}_1^\dagger\bd{D}_1^2\bd{\phi}_1 
+ \bd{\phi}_2^\dagger\bd{D}_1^2\bd{\phi}_2\right)},
\label{CF_f}
\end{eqnarray}
where 
$\bd{\phi}_1^\dagger\overset{\leftrightarrow}{\partial}_\mu 
\bd{\phi}_1
\equiv \bd{\phi}_1^\dagger{\partial}_\mu 
\bd{\phi}_1-({\partial}_\mu \bd{\phi}_1^\dagger) 
\bd{\phi}_1$.

\section{BPS equations}
\label{sc:BPSequation}

\subsection{ SUSY Vacuum}
\label{sc:susyvac}

For a stable BPS solution two or more isolated SUSY 
vacua are needed.
We investigate the vacuum structure of our model 
in this subsection.
The on-shell supertransformation for the fermionic 
fields are of the form:
\begin{eqnarray}
\delta_\varepsilon\psi^i_\mu &=&
{\cal D}_\mu\varepsilon^i - \frac{\kappa^2}{6}
\gamma_\mu M^I{\cal Y}_I{^i}{_j}\varepsilon^j,
\label{gravitino1}\\
\delta_\varepsilon \zeta^\alpha &=&
-\gamma^\mu{\cal D}_\mu{\cal A}_i{^\alpha}\varepsilon^i
- g_IM^I(t_I{\cal A}_i)^\alpha\varepsilon^i
+ \frac{\kappa^2}{2}{\cal A}_j{^\alpha}
M^I{\cal Y}_I{^j}{_i}\varepsilon^i,\label{hyperino1}
\end{eqnarray}
where ${\cal D}_\mu\varepsilon^i = \left(\partial_\mu 
- \frac{1}{4}\gamma_{ab}\omega_\mu{^{ab}}\right)
\varepsilon^i - \kappa^2V_{\mu}{^i}{_j}\varepsilon^j$,
${\cal D}_\mu{\cal A}_i{^\alpha} = \nabla_\mu
{\cal A}_i{^\alpha}
-\kappa^2V_{\mu i}{^j}{\cal A}_j{^\alpha}$ and 
$V_{\mu}{^{ij}} = - \frac{1}{2}{\cal
A}^{i\bar{\alpha}}\overset{\leftrightarrow}
{\partial}_\mu{\cal A}^j{_\alpha}$.
SUSY vacua conditions are obtained by
demanding `supersymmetric' spacetime independent 
background configuration for
scalar fields ${\cal A}_i{^\alpha}$ in
Eq.(\ref{hyperino1})\footnote{From Eq.(\ref{gravitino1}) 
a constraint 
for the killing spinor is obtained.}.
This yields the following condition:
\begin{eqnarray}
0 = \delta_\varepsilon\zeta^\alpha
\quad\Rightarrow\quad
(g_IM^It_I{\cal A}_i)^\alpha = \frac{\kappa^2}{2}
{\cal A}_j{^\alpha}M^0{\cal Y}_{0}{^j}{_i},
\end{eqnarray}
in the right hand side we used the constraint 
${\cal Y}_1=0$ in 
Eq.(\ref{2ndconst}). 
In terms of the Curtright-Freedman basis this can be
rewritten as 
\begin{eqnarray}
g_0\left(
\begin{array}{cc}
i\alpha & 0 \\
0 & -i\alpha\\
i\kappa\bd{D}\bd\phi_1 & i\kappa\bd{D}\bd\phi_2\\
i\kappa\bd{D}\bd\phi_2^* & - i\kappa\bd{D}\bd\phi_1^*
\end{array}
\right)
= -\frac{\kappa^2}{2}
\left(
\begin{array}{cc}
\mathcal{Y}_0{^{12}} & -\mathcal{Y}_0{^{11}}\\
\mathcal{Y}_0{^{22}} & -\mathcal{Y}_0{^{21}}\\
\kappa\left(\mathcal{Y}_0{^{12}}\bd\phi_1 
+ \mathcal{Y}_0{^{22}}\bd\phi_2\right) &
-\kappa\left(\mathcal{Y}_0{^{11}}\bd\phi_1 
+ \mathcal{Y}_0{^{21}}\bd\phi_2\right)\\
\kappa\left(\mathcal{Y}_0{^{22}}\bd\phi_1^* 
- \mathcal{Y}_0{^{12}}\bd\phi_2^*\right) &
-\kappa\left(\mathcal{Y}_0{^{21}}\bd\phi_1^* 
- \mathcal{Y}_0{^{11}}\bd\phi_2^*\right)
\end{array}
\right).
\end{eqnarray}
From the compensator components in the first and second rows, 
we find two conditions on the quantities defined in 
Eq.(\ref{eq:Ymatrix}) 
\begin{eqnarray}
{\cal T}_0 = \kappa^{-2}\alpha\bar{\cal A},\quad
{\cal S}_0 = 0.\label{vacuum_cond}
\end{eqnarray}
Using these conditions, we find a condition for $\phi_i{^a}$ 
as an eigenvalue equation for the matrix $\bd{D}$ :
\begin{eqnarray}
\bd{D}\left(
\begin{array}{c}
\bd\phi_1\\
\bd\phi_2
\end{array}
\right)
= \alpha
\left(
\begin{array}{c}
\bd\phi_1\\
-\bd\phi_2
\end{array}
\right).
\end{eqnarray}

For a generic choice of the parameters the diagonal 
matrix $\bd{D}$ should have 
no identical entries 
: $d_{a} \neq d_{b}$. 
Then the last condition yields $2(n+1)$ candidates 
for discrete SUSY vacua
such that the only one component of $\bd\phi_i$ is 
non zero and all 
the others components vanish. 
The vacuum expectation value (VEV) 
of the nonvanishing component 
$\phi_i{^a}$ in the $a$-th vacuum 
is determined through the equation $d_a = \pm
\alpha$ ($+$ is for $\bd\phi_{i=1}$ and $-$ is for 
$\bd\phi_{i=2}$).
More explicitly, this reduces to
\begin{eqnarray}
\left(\alpha_0d_{1,a}-d_{0,a}\alpha_1\right)
\left(d_{1,a}|\phi_i{^a}|^2 \mp 
\kappa^{-2}\alpha_1\right) = 0.
\end{eqnarray}
Therefore, the VEV at the $a$-th vacuum is given by 
\begin{eqnarray}
m_a
\equiv 
\frac{\kappa^{-2}\alpha_1}{d_{1,a}} 
\label{eq:VEV-a-vac}
\end{eqnarray}
Since $|\phi_i{^a}|^2 \ge 0$, the $a$-th vacuum 
with $a=1,2,\cdots,n+1$ is realized 
either by 
\begin{eqnarray}
|\phi_1{^b}|^2 &=& 
m_a\delta^b{_a},\ {\rm and}\ \bd\phi_2=0,
\label{eq:plus-vacuum}
\end{eqnarray}
in the case of $d_{1,a}\alpha_1 > 0$,  or by 
\begin{eqnarray}
|\phi_2{^b}|^2 &=& 
- m_a\delta^a{_b},\ {\rm and}\ \bd\phi_1=0 ,
\label{eq:minus-vacuum}
\end{eqnarray}
in the case of $d_{1,a}\alpha_1 < 0$. 
In what follows, we will choose without loss of 
generality\footnote{
The $U(1)$ gauge multiplet $W_\mu^1$ is introduced 
to obtain a curved target manifold. 
If the $U(1)$ charge of the $a$-th component of the 
hypermultiplet 
$d_{1,a}$ vanish, target space along that direction is flat 
(at least in the limit of vanishing gravitational coupling). 
We assume $d_{1,a}\neq 0$ in this article.}
$d_{1,a}\alpha_1>0$, which implies the 
$n+1$ isolated SUSY vacua given in 
Eq.(\ref{eq:plus-vacuum}). 
Note that these SUSY vacua satisfy the constraint 
${\cal T}_1 = {\cal S}_1 =0$ in Eq.(\ref{2ndconst_2}) and remaining vacuum condition 
${\cal T}_0 =\kappa^{-2}\alpha\bar{\cal A}$ and 
${\cal S}_0 = 0$ in Eq.(\ref{vacuum_cond}).
In order for these VEV's $m_a$ defined in 
Eq.(\ref{eq:VEV-a-vac}) 
to be finite in the 
weak gravity limit, 
we must require $\alpha_1$ to be proportional to 
$\kappa^2$ in the limit of vanishing gravitational coupling 
$\kappa \rightarrow 0$. 

\begin{figure}[t]
\begin{center}
\includegraphics{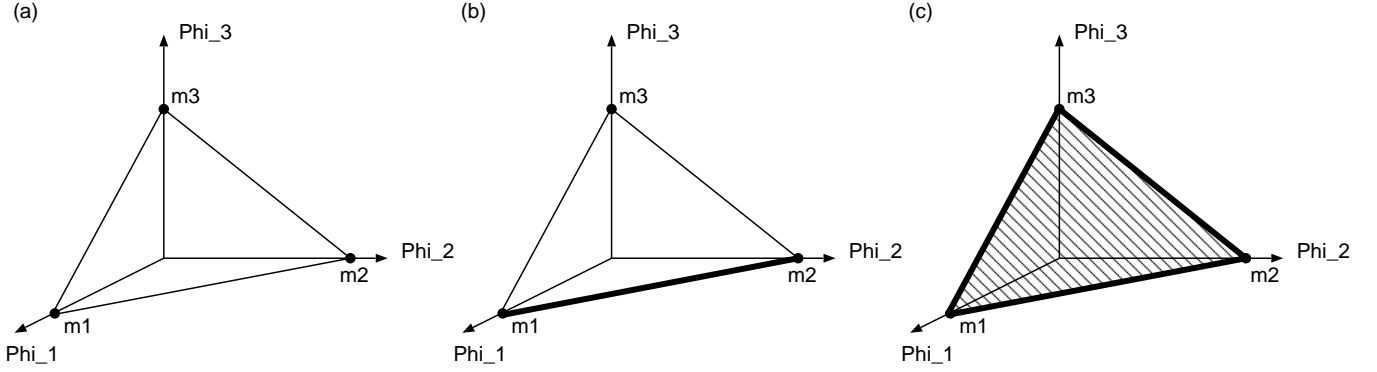}
\caption{The SUSY vacuum structure for $n=2$ is shown. 
Three isolated vacua 1,2,and 3 are shown in (a). 
An isolated vacuum 3 and a (1, 2) degenerate vacuum 
are shown in (b). 
A (1, 2, 3) degenerate vacuum is shown in (c).}
 \label{fig_vacuum}
\end{center}
\end{figure}

It is useful to introduce a geometrical picture of 
field space in order to clarify the structure of 
these SUSY vacuum. 
We make the ansatz 
\begin{equation}
\bd{\phi}_2=0 , 
\label{eq:vac-phi2-ansatz}
\end{equation}
assuming $d_{1,a}\alpha_1>0$. 
Then the one 
of the constraint ${\cal S}_1=0$ is automatically 
satisfied. 
Another constraint ${\cal T}_1= 0$ is most conveniently 
expressed by using modulus squared variables  
\begin{eqnarray}
\Phi^a \equiv |\phi_1{^a}|^2, 
\label{eq:modulus-squared}
\end{eqnarray}
\begin{eqnarray}
1 = \sum_{b=1}^{n+1} \frac{\Phi^b}{m_b}. 
\label{T_1=0}
\end{eqnarray}
This means that the field space is restricted to the 
non-negative portion of an $n$ dimensional 
hyperplane in the $n+1$ dimensional space. 
We call this hyperplane $\Sigma_1^{(n)}$. Namely,
$\Sigma_1^{(n)}$ is an $n$ simplex 
:
\begin{eqnarray}
\Sigma_1^{(n)} = \left\{\bd{\Phi}\in\mathbb{R}^n\ \bigg|\ 
\bd{\Phi} = \sum_{b=1}^{n+1}c_b\bd{e}_b,\quad c_b\ge0,
\quad \sum_{b=1}^{n+1}{c_b \over m_b}=1\right\},
\label{eq:n-simplex}
\end{eqnarray}
where $\bd{e}_a$ denotes a unit vector.
 For the case of generic parameters 
$(d_{a}\neq d_{b})$, there are $n+1$ isolated 
vacua, each of which corresponds to a 
vertex of the $n$ simplex.

Now we turn our attention to the case of degenerate vacua. 
We always assume $\bd\phi_2=0$ as in 
Eq.(\ref{eq:vac-phi2-ansatz}) and use 
${\Phi}^a$ in Eq.(\ref{eq:modulus-squared}) as our variables. 
Some of the above $n+1$ SUSY vacua are degenerate 
when the parameters of our model 
take particular values.
Let us work out the case of $\ell$ eigenvalues 
$d_a\equiv d_{0,a}+fd_{1,a}$ of 
the matrix $\bd{D}$ are degenerate. 
Without loss of generality, we assume that the first 
$l$ eigenvalues $a=1, \cdots, \ell$ are degenerate: 
\begin{equation}
\alpha_0 + f\alpha_1 = d_{0,1} + fd_{1,1} = \cdots 
= d_{0,\ell} + fd_{1,\ell}, 
\label{dege_vac}
\end{equation}
which imply the following condition among 
the parameters of the $\ell$ degenerate vacua 
\begin{eqnarray}
\frac{\alpha_0 - d_{0,1}}{\alpha_1 - d_{1,1}}
=\frac{\alpha_0 - d_{0,2}}{\alpha_1 - d_{1,2}}
=\cdots =
\frac{\alpha_0 - d_{0,\ell}}{\alpha_1 - d_{1,\ell}}
.
\label{eq:degenerate-vac-cond}
\end{eqnarray}
Since other eigenvalues are not degenerate with these, 
the eigenvalue condition implies $\Phi^a=0$ for 
$a \not\in (1, \cdots, \ell)$. 
The value of the ratio in 
Eq.(\ref{eq:degenerate-vac-cond}) is precisely 
$-f(\phi)$ defined in Eq.(\ref{CF_f}) : 
\begin{eqnarray}
- f = \frac{\displaystyle \alpha_0 - \sum_{b=1}^\ell
d_{0,b}\dfrac{\Phi^b}{m_b}}{
\displaystyle \alpha_1 - \sum_{b=1}^\ell
d_{1,b}\dfrac{\Phi^b}{m_b}
}.
\end{eqnarray}
Plugging $d_{0,a} = -fd_{1,a} + \alpha_0 + f\alpha_1$ 
into this, we find 
$\displaystyle
(\alpha_0 + f\alpha_1)
\left(1 - \sum_{b=1}^\ell\dfrac{\Phi^b}{m_b}\right) = 0
$ 
giving a condition 
for the fields in the degenerate 
vacua : 
\begin{equation}
\sum_{a=1}^\ell {{\Phi}^a \over m_a}=1.
\label{eq:constr-deg-vac}
\end{equation}
Therefore, we find the degenerate SUSY vacuum is given by 
the positive portion of an $\ell-1$ dimensional 
hypersurface defined by (\ref{eq:constr-deg-vac}) 
with $\ell$ moduli, 
in contrast to the discrete vertex points of the 
$n$-dimensional simplex $\Sigma_1^{(n)}$ in the case of 
nondegenerate SUSY vacua. 
Notice that this is consistent in the
constraint ${\cal T}_1\big| =0$.

Clearly the degenerate vacuum becomes a subsimplex 
$\sigma_1^{(l)}$ embedded in the $n$ simplex 
$\Sigma_1^{(n)}$ 
of the entire field space. 
In terms of a convenient parameter for following 
discussion defined as 
\begin{equation}
\nu_{I,a} \equiv 2g_0M^0\left(\alpha_I- d_{I,a}\right), 
\label{def_nu}
\end{equation}
the condition of the degeneracy of $a$ and $b$ vacua 
is given by 
\begin{equation}
{\nu_{0,a} \over \nu_{1,a}}
=\cdots=
{\nu_{0,b} \over \nu_{1,b}}
. 
\end{equation}
Whenever the ratio of parameters 
$\nu_{0,a}/\nu_{1,a}$ 
coincide for different $a$, their vacua are degenerate. 
All the remaining vacua are 
isolated and its structure is the same as the general 
nondegenerate case.
If $\ell(<n)$ vacua are degenerate, there is a unique 
 $\ell$-dimensional subsimplex $\sigma^{(\ell)}_1$ 
 as a continuum of degenerate SUSY vacua. 
This entire subsimplex itself 
becomes a continuum of the SUSY vacua in field space. 
We illustrate this situation in
Fig.\ref{fig_vacuum}. 
In the maximally degenerate case, 
the entire portion of the hyperplane $\Sigma^{(n)}_1$ 
corresponds to the SUSY vacua, 
so that there is no stable 
domain wall solution which interpolates 
between two isolated vacua. 

\subsection{ BPS equations}

Instead of solving Einstein equations directly, 
it is easy to obtain classical solutions by solving 
the BPS equations conserving a half of SUSY.
In this subsection we first derive the BPS equations for 
the hypermultiplet scalar fields ${\cal A}_i{^\alpha}$ 
and we rewrite it in terms of the fields 
$\phi_i{^a}$ in the Curtright-Freedman basis 
(\ref{CFbasis}). 
We finally obtain the BPS
equations for the modulus squared fields $\Phi^a$ 
defined in Eq.(\ref{eq:modulus-squared}), 
using a certain ansatz, which will be justified a 
posteriori.

If we assume the warped metric (\ref{5dmetric}),  
the SUSY transformation of gravitino (\ref{gravitino1}) 
decouples into two parts 
\begin{eqnarray} 
  \delta_\varepsilon \psi_m^i 
   &=& \partial_m \varepsilon^i 
       - \frac{1}{2}\gamma_m \gamma^y 
       \partial_y U\cdot \varepsilon^i 
       - \frac{\kappa^2}{6}M^I{\cal Y}_I{^i}{_j}
         \gamma_m \varepsilon^j ,
  \label{gravitino2a} \\
  \delta_\varepsilon \psi_y^i 
   &=& \partial_y \varepsilon^i - \kappa^2 V_y{}^i{}_j 
   \varepsilon^j
        - \frac{\kappa^2}{6}M^0{\cal Y}_0{}^i{}_j
          \gamma_y \varepsilon^j .
  \label{gravitino2b}
\end{eqnarray}

Let us require vanishing of the SUSY variation of 
gravitino and hyperino 
to preserve the following four SUSY (out of eight SUSY) 
specified by 
\begin{eqnarray}
\gamma^y \varepsilon^i(y)= i \tau_3{}^{i}{}_j \varepsilon^j(y),
\label{1/2SUSY}
\end{eqnarray}
where $\tau_3$ is one of the Pauli matrix.
Then one of the gravitino BPS conditions (\ref{gravitino2a}) 
gives an equation for the function $U(y)$ in 
the warp factor and an 
additional constraint 
\begin{eqnarray}
 \partial_y U &=& i\frac{\kappa^2}{3}M^I{\cal Y}_I{^1}{_1},
\label{BPS-warp}\\
0 &=& M^I{\cal Y}_I{^1}{_2}.
\label{add-const}
\end{eqnarray}
The hyperino BPS condition (\ref{hyperino1}) gives 
\begin{eqnarray}
\nabla_y\mathcal{A}_1{^\alpha} 
= - \kappa^2V_{y1}{^j}\mathcal{A}_j{^\alpha}
+ ig_IM^I(t_I\mathcal{A}_1)^\alpha
- i \frac{\kappa^2}{2}\mathcal{A}_1{^\alpha}M^I
\mathcal{Y}_{I}{^1}{_1}\label{BPS_hype}
\end{eqnarray}
Since Eq.(\ref{1/2SUSY}) assures that solutions of these BPS 
equations conserve four SUSY out of eight SUSY, the effective theory on 
this background has ${\cal N}=1$ SUSY in four dimensions. 
This should be useful for 
model building in the 
SUSY brane-world scenario. 

Let us turn to rewrite these BPS equations in terms of 
$\phi_i{^a}$.
Since we are interested in the kink solutions 
which interpolate the two
SUSY vacua in Eq.(\ref{eq:plus-vacuum}), we take an ansatz 
that all the $\bd\phi_2$ vanish by 
considering the case where $d_{1,a}\alpha_1>0$ 
\begin{eqnarray}
\bd{\phi}_2 = 0. 
\label{eq:phi2-ansatz}
\end{eqnarray}
The BPS equation 
for the function $U$ in the 
warp factor (\ref{BPS-warp}) reduces to
\begin{eqnarray}
\partial_y U = 
- \frac{2}{3}\kappa^2\bar{\cal A}^{-1}g_IM^I{\cal T}_I.
\label{BPS-warp2}
\end{eqnarray}
The additional constraint (\ref{add-const}) reduces 
to $g_IM^I{\cal S}_I = 0$, which is automatically 
satisfied by the ansatz (\ref{eq:phi2-ansatz}).
We also take an 
ansatz 
\begin{eqnarray}
\phi_1{^a} = {\rm e}^{i\theta_a}\times \mathbb{R}
\label{eq:ansatz-phi1}
\end{eqnarray}
 where $\theta_a$ is a constant phase. 
These ansatz (\ref{eq:phi2-ansatz}) and 
(\ref{eq:ansatz-phi1}) give $V_y = W^1_y=0$. 
Therefore, Eq.(\ref{BPS_hype}) reduces to
\begin{eqnarray}
\partial_y {\cal A}_1{^\alpha} = 
ig_IM^I(t_I{\cal A}_1)^\alpha 
+ \kappa^2\bar{\cal A}^{-1}g_IM^I
{\cal T}_I{\cal A}_1{^\alpha}.
\end{eqnarray}
Since we choose $t_I$ as diagonal matrices, 
this reduces to $2 + 2(n+1)$ 
decoupled equations. 
The second component ($\alpha = 2$) and 
last $n+1$ components ($\alpha \ge n+4$) 
are automatically satisfied. 
For $\alpha = 1$ this takes the form:
\begin{eqnarray}
\partial_y \bar{\cal A}^{-\frac{1}{2}}
= - g_0M^0\bar{\cal A}^{-\frac{1}{2}}
\left( \alpha 
- \kappa^2\bar{\cal A}^{-1}{\cal T}\right),
\label{BPS_add}
\end{eqnarray}
and equations for $\phi_1{^a}$ can be derived from 
$\alpha = a+2=3,4,\cdots,n+3$:
\begin{eqnarray}
\partial_y\left(\kappa^{-1}\bar{\cal A}^{-\frac{1}{2}}
\phi_1{^a}\right)
= - g_0M^0\left(d_a - \kappa^2\bar{\cal A}^{-1}
{\cal T}\right)
\left(\kappa^{-1}\bar{\cal A}^{-\frac{1}{2}}
\phi_1{^a}\right),
\label{BPS_phi}
\end{eqnarray}
where ${\cal T} \equiv {\cal T}_0 + f{\cal T}_1$.

In order to obtain BPS equations for $\Phi^a$ 
defined in Eq.(\ref{eq:modulus-squared}), we multiply
$\phi_1{^{a*}}$ to both side of the Eq.(\ref{BPS_phi}):
\begin{eqnarray}
\sum_{b=1}^{n+1}\bd{K}^a{_b}\partial_y\Phi^b 
= - 2g_0M^0\left(\bar{\cal A}d_a - \kappa^2{\cal T}\right)
\Phi^a, \label{BPS_phi2}
\end{eqnarray}
where 
$\bd{K}^a{_b} = \bar{\cal A}\delta^a{_b} + \kappa^2\Phi^a$ 
and $(\bd{K}^{-1})^a{_b} = \bar{\cal A}^{-1}(\delta^a{_b} -
\kappa^2\Phi^a)$. 
The lack of column indices $b$ in the second 
term of the right hand side implies 
that it appears in each column of the matrices.
Eq.(\ref{BPS_phi2}) can be 
rewritten for $a= 1, \cdots, n+1$ 
\begin{eqnarray}
\partial_y\Phi^a &=& 
-2g_0M^0\left[(d_{0,a}-\alpha_0) 
+ f (d_{1,a} - \alpha_1)\right]\Phi^a,
\label{BPS_phi3}\\
f &=& - \frac{\displaystyle \kappa^{-2}\alpha_0\alpha_1 
- \sum_{b=1}^{n+1}d_{0,b}d_{1,b}\Phi^b}{\displaystyle 
\kappa^{-2}\alpha_1^2 - \sum_{b=1}^{n+1}d_{1,b}^2\Phi^b},
\end{eqnarray}
where the following identities for $\bd{K}^{-1}$ are used 
\begin{eqnarray}
\sum_{b=1}^{n+1}(\bd{K}^{-1})^a{_b}\Phi^b = \Phi^a,\quad
\bar{\cal A}\sum_{b=1}^{n+1}(\bd{K}^{-1})^a{_b}d_b\Phi^b
= (d_a-\alpha+\kappa^2{\cal T})\Phi^a.
\end{eqnarray}
All the SUSY vacua 
$\Phi^b =m_a\delta^b{_a} \quad (a=1,2,\cdots,n+1)$
correspond to fixed points of these BPS equations 
(\ref{BPS_phi3}). 
Solutions of this BPS equations must satisfy the 
constraint\footnote{
Another constraint ${\cal S}_1 = {\cal S}_1^* = 0$ is 
automatically satisfied by the ansatz $\bd\phi_2=0$ in 
Eq.(\ref{eq:phi2-ansatz}).} 
${\cal T}_1=0$ in Eq.(\ref{T_1=0}). 
We find that the solutions with an initial condition 
${\cal T}_1=0$, say at $y=y_0$, satisfy the constraint 
at any $y$. 
We can also show that the additional BPS equation 
(\ref{BPS_add}) is automatically 
satisfied by solutions of Eq.(\ref{BPS_phi3}).

We can eliminate $\Phi^{n+1}$ through 
the constraint ${\cal T}_1=0$ in Eq.(\ref{T_1=0}) 
and transform the above BPS equations for $n+1$ 
(dependent) fields into the BPS equations for 
$n$ (independent) fields 
by letting $a$ to run $1$ to $n$ in Eq.(\ref{BPS_phi3})
and by rewriting $f$ as 
\begin{eqnarray}
f = - \frac{\displaystyle \kappa^{-2}\alpha_1(\alpha_0 - d_{0,n+1}) 
- \sum_{b=1}^n(d_{0,b} - d_{0,n+1})d_{1,b}\Phi^b}{\displaystyle 
\kappa^{-2}\alpha_1(\alpha_1 - d_{1,n+1}) 
- \sum_{b=1}^n(d_{1,b} - d_{1,n+1})d_{1,b}\Phi^b}.
\end{eqnarray}
We can always change the $n+1$ fields formalism to the $n$ 
fields formalism and vice versa 
according to our purpose.

After checking the consistency between the BPS equations 
and the
constraint ${\cal T}_1 = 0$, the BPS equations can be 
simplified 
\begin{eqnarray}
\partial_y \Phi^a = 
\left[\nu_{0,a} + \nu_{1,a}f(\Phi)\right]\Phi^a,\quad
f(\Phi) = -\dfrac{\displaystyle \sum_{b=1}^{n+1}\nu_{0,b}
\dfrac{\Phi^b}{m_b}}{\displaystyle \sum_{b=1}^{n+1}\nu_{1,b}
\dfrac{\Phi^b}{m_b}},
\qquad 
a=1,2,\cdots,n+1, 
\label{BPS_Phi}
\end{eqnarray}
where $\nu_{I,a}$ is defined in Eq.(\ref{def_nu}).
This is the final form of our BPS equations.

The scalar potential 
(\ref{CFpot}) for the section $\bd \phi_2=0$ of field space 
is expressed in terms of $\Phi^a$ as\footnote{
We give a more compact form of this scalar potential 
in Appendix B, see
Eq.(\ref{pot_simple}) or Eq.(\ref{pot_suppot}).
}
\begin{eqnarray}
V(\Phi) = \frac{1}{2}\bar{\cal A}^{-1}\left[
\sum_{b=1}^{n+1}\nu_{0,b}^2 \Phi^b 
- \dfrac{\displaystyle \left(\sum_{b=1}^{n+1}\nu_{0,b}
\dfrac{\Phi^b}{m_b}\right)^2}{\displaystyle
\sum_{b=1}^{n+1}\dfrac{1-\kappa^2m_b}{m_b}
\dfrac{\Phi^b}{m_b}}
+ \kappa^2
\frac{\displaystyle \left(\sum_{b=1}^{n+1}\nu_{0,b}
\Phi^b\right)^2}{
\displaystyle \sum_{b=1}^{n+1}
\dfrac{1-\kappa^2m_b}{m_b}\Phi^b}
\right]
 - \frac{2\kappa^2}{3}
\left(a + \frac{\displaystyle 
\sum_{b=1}^{n+1}\nu_{0,b}\Phi^b}{
\displaystyle \sum_{b=1}^{n+1}
\dfrac{1-\kappa^2m_b}{m_b}\Phi^b}\right)^2,
\label{scalar_pot}
\end{eqnarray}
where we define 
\begin{eqnarray}
a \equiv 2g_0M^0\kappa^{-2}\alpha_0.
\label{eq:def-a}
\end{eqnarray}
The first term of the right-hand side of Eq.(\ref{scalar_pot}) 
vanishes at the SUSY vacua. 
The vacuum energy at each vacuum is determined by 
the second term. 
The vacuum energy density at 
$\Phi^b = m_a\delta^b{_a}$ is non-positive definite 
and is suppressed by the gravitational coupling
$\kappa^2$:
\begin{eqnarray}
V_{{\rm vacuum} \, a} = - \frac{2\kappa^2}{3}
\left(a + \frac{m_a}{1-\kappa^2m_a}\ \nu_{0,a}\right)^2.
\end{eqnarray}
This means that the SUSY vacuum becomes $AdS_5$ spacetime 
for general choice 
of the parameters and a flat spacetime for a special 
case where 
$d_{0,a}m_a - \kappa^{-2}\alpha_0 = 0$.

\begin{figure}[t]
\includegraphics{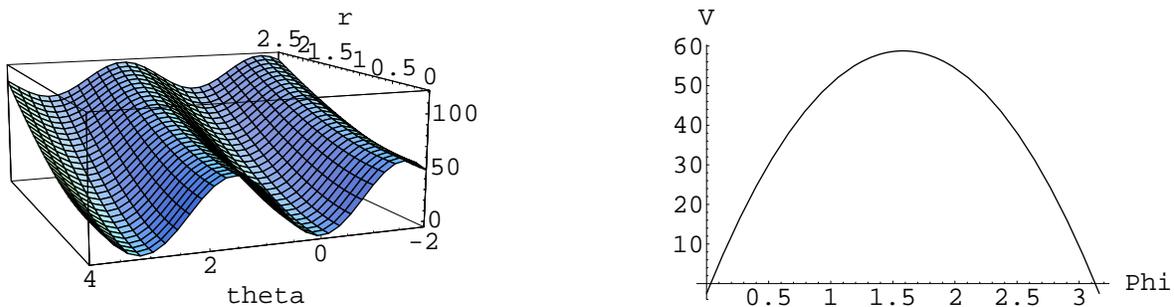}
\caption{Scalar potential $V$ of $T^*{\bf C}P^1$ model 
exhibiting discrete vacua. 
Left figure gives the potential as a function of 
two independent variables $r, \theta$. 
Vacua are located at $(r, \theta)=(0, 0), \, (0, \pi)$. 
The BPS wall solution interpolates these two vacua through 
the path $r=0$. 
Right figure shows the potential along the $r=0$ section 
as a function of $\Phi=\Phi^1$. 
Parameters are taken to be $(g_0M^0, \kappa) = (3, 0.1)$,
 $(\alpha_1,d_{1,1},d_{1,2})=(10^{-\frac{3}{2}},1,1)$ and
 $(\alpha_0,d_{0,1},d_{0,2}) = (0,-1,1)$.}
\label{pot-plot}
\end{figure}

Notice that the scalar potential (\ref{scalar_pot}) covers 
only a part of the field space, since we took the 
ansatz $\phi_1{^a} = {\rm e}^{i\theta_a}\times \mathbb{R}$ 
in Eq.(\ref{eq:ansatz-phi1}) 
and $\bd{\phi}_2 =0$ in Eq.(\ref{eq:phi2-ansatz}). 
However, it covers completely the section of field space 
taken by the field configuration of the BPS multi-wall 
solutions. 
To illustrate the situation, we take $n=1$ case with the 
symmetric kinetic term 
$d_{1,1}=d_{1,2}=1$ and moreover 
$d_{0,1}=-d_{0,2}=-1$. 
Previously we have constructed a coordinate 
which satisfies the 
constraint ${\cal S}_1 = {\cal S}_1^* = 0$ and 
${\cal T}_1=0$ and completely 
covers all the region of the target manifold 
\cite{ANNS}, \cite{ANNS2}:
\begin{eqnarray}
\left(
\begin{array}{cc}
\phi_1{^1} & \phi_1{^2}\\
\phi_2{^1} & \phi_2{^2}
\end{array}
\right)
= \left(
\begin{array}{cc}
g(r) {\rm e}^{\frac{i}{2}(\Psi + \Theta)} 
\cos \frac{\theta}{2}  &
g(r) {\rm e}^{\frac{i}{2}(\Psi - \Theta)} 
\sin \frac{\theta}{2} \\
f(r) {\rm e}^{-\frac{i}{2}(\Psi - \Theta)} 
\sin \frac{\theta}{2} &
-f(r) {\rm e}^{-\frac{i}{2}(\Psi + \Theta)} 
\cos \frac{\theta}{2} 
\end{array}
\right),
\end{eqnarray}
where we set
\begin{eqnarray}
f(r)^2 = \frac{1}{2}\left(-\kappa^{-2}\alpha_1 
+ \sqrt{4r^2 + \kappa^{-4}\alpha_1^2}\right),\quad
g(r)^2 = \frac{1}{2}\left(\kappa^{-2}\alpha_1 
+ \sqrt{4r^2 + \kappa^{-4}\alpha_1^2}\right).
\end{eqnarray}
The scalar potential is shown as a function of $r$ and 
$\theta$ in Fig.\ref{pot-plot}. 
The SUSY vacua become local minima of the scalar potential.
BPS domain wall which interpolates between these 
two vacua $(r,\theta) =
(0,0)$ and $(0,\pi)$ was obtained as \cite{ANNS}, 
\cite{ANNS2}
\begin{eqnarray}
r=0,\quad \cos\theta = \tanh(2g_0M^0(y-y_0)),\quad 
\Theta = {\rm const.}, 
\end{eqnarray}
and $\Psi$ undetermined. 
Hence, the wall trajectory runs only on the section $r=0$.
Let us turn to our ansatz.
Our ansatz (\ref{eq:phi2-ansatz}) and 
(\ref{eq:ansatz-phi1}) correspond to the section : 
$r=0$, $\Theta = 2\theta_1$ and 
$\Psi$ undetermined. 
Therefore our ansatz precisely covers the field 
space traversed by the domain wall configuration 
interpolating the two vacua as illustrated in 
Fig.\ref{pot-plot}.

We note also a simplification for the special case 
where we choose the symmetric kinetic term 
$d_{1,a}=d_{1,n+1}$ for the $U(1)$ generator $t_1$ 
as in Eq.(\ref{eq:sym-kin-term}). 
The BPS equations (\ref{BPS_Phi}) becomes very simple:
\begin{eqnarray}
\partial_y\Phi^a = \left[\nu_{0,a} - \frac{1}{m}
\sum_{b=1}^{n+1}\nu_{0,b}\Phi^b\right]\Phi^a\quad
(a=1,2,\cdots,n+1), \label{diagonal_U(1)}
\end{eqnarray}
and the VEV reduces to $\Phi^b = m\delta^b{_a}$. 
This is the case which was studied extensively 
in the global SUSY case ($\kappa\rightarrow 0$) 
previously \cite{GTT}. 
The scalar potential also becomes simple:
\begin{eqnarray}
V = \frac{1}{2(1-\kappa^2m)}\left[
\sum_{b=1}^{n+1}\nu_{0,b}^2\Phi^b - 
\frac{1}{m}\left(\sum_{b=1}^{n+1}
\nu_{0,b}\Phi^b\right)^2\right]
- \frac{2\kappa^2}{3}\left(
a + \frac{1}{1-\kappa^2m}
\sum_{b=1}^{n+1}\nu_{0,b}\Phi^b\right)^2.
\end{eqnarray}
The first term is precisely the same as the potential 
in the global SUSY case except a gravitational correction 
of the overall factor, whereas 
the second term gives a genuine gravitational correction.
We will study this potential more closely 
when discussing 
the weak gravity limit in sect.\ref{sc:discuss}.

\section{Exact BPS solutions}
\label{sc:exactBPSsol}

\subsection{General properties}

As we showed in sect.\ref{sc:susyvac}, 
the field space of our BPS solutions 
is restricted in the $n$ simplex $\Sigma_1^{(n)}$ 
in Eq.(\ref{eq:n-simplex}) 
with the SUSY vacua as their $n+1$ vertices 
(or subsimplexes for degenerate cases). 
Our solutions of the BPS equation will be expressed 
as one-dimensional curves which interpolates between 
two vertices (or subsimplexes for degenerate cases) 
of $\Sigma^{(n)}_1$.
These curves never intersect each other, since these are the solutions
of the first order ordinary differential equations.

Let us consider a BPS kink interpolating between two SUSY 
vacua $c$ and $d$ : 
\begin{eqnarray}
\begin{array}{l}
\Phi^a \rightarrow m_a\delta^a{_c}\quad{\rm as}
\quad y \rightarrow - \infty,\\
\Phi^a \rightarrow m_a\delta^a{_d}
\quad{\rm as}\quad y \rightarrow + \infty.
\end{array}\label{gene_boundary_cond}
\end{eqnarray}
Since the variables $\Phi^a$ satisfy $\sum_a \Phi^a/m_a=1$ 
as given in Eq.(\ref{T_1=0}) and 
$\Phi^a\ge 0$ by definition, $\Phi^{a\neq c,d}$ must 
vanish as $y \rightarrow \pm \infty$. 
Near the two vacua (\ref{gene_boundary_cond}), 
the BPS equations (\ref{BPS_Phi}) behave as 
\begin{eqnarray}
\begin{array}{l}
\partial_y\Phi^a \sim \nu_{1,a}
\left(\dfrac{\nu_{0,a}}{\nu_{1,a}} 
- \dfrac{\nu_{0,c}}{\nu_{1,c}}\right)\Phi^a
\quad {\rm as}\ y\rightarrow-\infty,\\
\partial_y\Phi^a \sim \nu_{1,a}
\left(\dfrac{\nu_{0,a}}{\nu_{1,a}} 
- \dfrac{\nu_{0,d}}{\nu_{1,d}}\right)\Phi^a
\quad {\rm as}\ y\rightarrow+\infty.
\end{array}
\label{eq:asympt-BPS-asym}
\end{eqnarray}
To find out generic features of BPS wall solutions, 
it is useful to determine ordering of vacua \cite{GTT}. 
For general choices of $\alpha_I$ and $d_{I,a}$ it is 
a complicated matter to 
determine the ordering of the SUSY vacua. 
 For simplicity, we assume that 
$\nu_{1,a} = 2g_0M^0\left(\alpha_1 - d_{1,a}\right) < 0$ 
for any $a$. 
This assumption\footnote{
In the case of the symmetric kinetic term, this assumption is not needed
as we see below. 
} is always realized in the 
weak gravity limit 
where $\alpha_1\sim {\cal O}(\kappa^2) \ll 1$. 
The above asymptotic behavior of the BPS equations 
(\ref{eq:asympt-BPS-asym}) suggests that the ordering 
of vacua should be made with respect to the ratio 
$\nu_{0,a}/\nu_{1,a}$. 
Here we assume the ordering of vacua as 
\begin{equation}
\dfrac{\nu_{0,1}}{\nu_{1,1}} > \cdots >
\dfrac{\nu_{0,n+1}}{\nu_{1,n+1}}. 
\label{eq:vac-ordering}
\end{equation}
Let us consider the case $c<d$. 
Combining the asymptotic form (\ref{eq:asympt-BPS-asym}) 
and the above boundary conditions 
(\ref{gene_boundary_cond}), we find that  
$\Phi^a$ with $a<c$ and $a>d$ must vanish identically 
at any $y$. 
Therefore we need to consider fields $\Phi^a$ for 
$c\le a \le d$. 
Since $\Phi^a$ with $c<a<d$ must vanish as 
$y\rightarrow \pm \infty$, they have at least 
one maximum.

For simplicity, we consider the case of symmetric 
kinetic term $d_{1,a} = d_{1,n+1}$, $a=1, \cdots, n$ 
as in Eq.(\ref{eq:sym-kin-term}). 
Ordering of vacua should now be specified by 
$\nu_{0,1} < \cdots < \nu_{0,n+1}$.
If we consider a kink interpolating between $c$-th 
and $d$-th vacua, we have again nonvanishing fields 
$\Phi^a, a=c, \cdots, d$ which now satisfies a 
simplified BPS equations 
\begin{eqnarray}
\partial_y\Phi^a = 
\frac{1}{m}\left(\sum_{b=c}^{d}
(\nu_{0,a} -\nu_{0,b})\Phi^b\right)\Phi^a, 
\quad\quad \left(c\le a \le d\right).
\label{eq:sym-kin-BPS-eq}
\end{eqnarray}
This BPS equation implies that 
$\Phi^{a=c,d}$ is monotonic functions of $y$. 
Then we find that the curves which interpolate 
between $c$-th and 
$d$-th vacua is embedded in the $\ell = d-c$ 
subsimplex $\sigma^{(l)}_1$ in $\Sigma^{(n)}_1$. 
Therefore it is enough to consider solutions 
which interpolate between the first and the last vacuum. 
This embedding of multi-wall solutions in the subsimplexes 
is an extension of a feature already noted in the global 
SUSY model with the symmetric kinetic term in Ref.\cite{GTT}.

\begin{figure}[t]
\begin{center}
\includegraphics{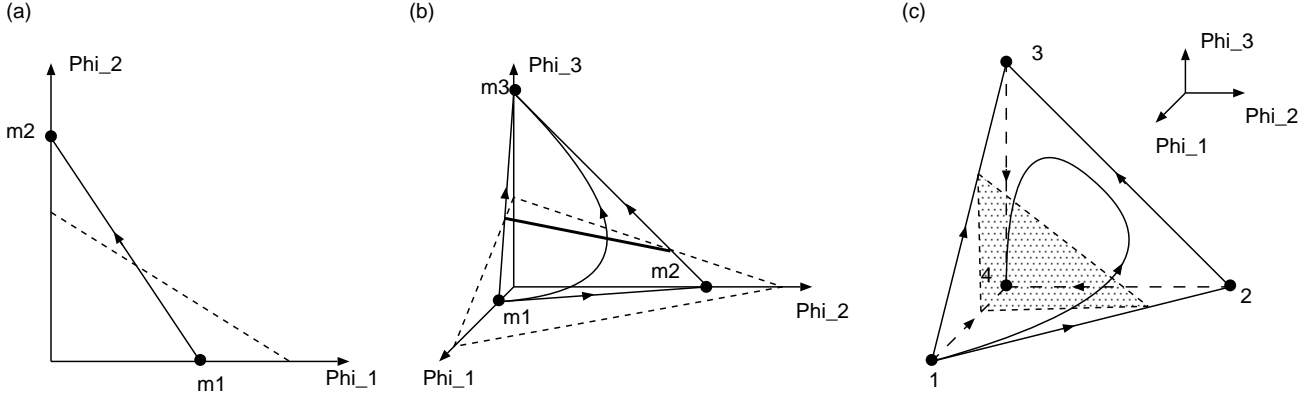}
\caption{Solution space for $n=1,2,3$. Solid lines express the simplex
 $\Sigma_1^{(n)}$ and broken lines denote $\Sigma_0^{(n)}$. The curves
 in $\Sigma_1^{(n)}$ denotes the solution curve.}\label{example}
\end{center}
\end{figure}

One of the advantages of using 
the geometrical picture in terms of $\Phi^a$ 
is that we can obtain the 
information about the warp factor very easily. 
In four-dimensional SUGRA 
\cite{CQR,EMSS} or the $D$-dimensional 
effective SUGRA 
\cite{Cvetic-Lam,Skenderis:1999mm}, 
the killing spinor
equations are obtained:
\begin{eqnarray}
\partial_y\phi^i(y) = C_D 
\frac{\partial W(\phi^i)}{\partial\phi^i},\quad
\partial_yU(y) = - \kappa^2C'_D W(\phi^i),
\label{eq:killing-spinor-eq}
\end{eqnarray}
where $C_D$ and $C'_D$ are positive constants which 
depend on spacetime dimension $D$. 
A real function $W$ is called ``superpotential''.
Therefore, necessary and sufficient condition 
for the presence of 
the stationary point of the warp factor 
is given by the existence of zero of the 
``superpotential'' along the wall 
trajectory. 
From Eq.(\ref{eq:killing-spinor-eq}) we obtain 
\begin{eqnarray}
\partial_y^2U 
= -\kappa^2C_DC'_D\sum_i\left(\dfrac{\partial W}
{\partial \phi^i}\right)^2 \le 0 .
\end{eqnarray}
Since $\partial_y^2 U$ is non-positive definite, 
$\partial_yU$ is always a monotonic function, and 
the warp factor has at most a single stationary point.
Eq.(\ref{eq:killing-spinor-eq}) shows that 
the warp factor can vanish at both infinity resulting in 
a space-time regular in the entire bulk, 
if the superpotential $W$ 
has odd numbers of zeroes along the wall trajectory. 
On the contrary, if the superpotential has no 
zero at all (or even numbers of zeros), 
the warp factor is singular 
at least at either one side of the wall. 
In that case, the BPS wall solution expresses the 
coupling flow of certain four-dimensional 
field theory according to the 
AdS/CFT correspondence. 

Let us introduce ``superpotential'' for our model.
We define it as a derivative of the function $U$ in the 
warp factor 
in Eq.(\ref{BPS-warp2}):
\begin{eqnarray}
\partial_y U = 
-\frac{\kappa^2}{3}(2g_0M^0\bar{\cal A}^{-1}{\cal T}_0) 
\equiv -\frac{\kappa^2}{3}W.
\label{BPS_warp_superpot}
\end{eqnarray}
Here, we impose the constraint ${\cal T}_1=0$. 
We will discuss the superpotential $W$ in some detail 
in Appendix B.
The warp factor has a stationary point at a point where $W=0$.
The condition $W=0$ leads to ${\cal T}_0=0$ from its 
definition (\ref{BPS_warp_superpot}).
This ${\cal T}_0 = 0$ condition defines a 
hyperplane which we shall 
call  $\Sigma_0^{(n)}$. 
We also denote the intersection of 
$\Sigma_1^{(n)}$ and $\Sigma_0^{(n)}$ 
as $\Omega^{(n-1)}$.
The $\partial_yU$ vanishes on the intersection 
$\Omega^{(n-1)}$. 
This means that the warp factor does not have
a stationary point when $\Sigma_1^{(n)}$ and 
$\Sigma_0^{(n)}$ do not
intersect each other. 
On the other hand, there exist some stationary 
points of $W$, if 
the $\Sigma_0^{(n)}$ and $\Sigma_1^{(n)}$ intersect.
The number of the stationary points of $W$ equals to 
the number of intersections between the trajectory 
and $\Omega^{(n-1)}$.
The warp factor increases (decreases) in the lower (upper) 
half space
below (above) the hyperplane $\Sigma^{(n)}_0$ along the 
wall trajectory.
We show this circumstance in Fig.\ref{example} for $n=1,2,3$. 
At this stage we have a question whether the wall 
trajectory can have
two or more intersections with $\Omega^{(n-1)}$.
It is familiar that $\partial_y^2U<0$ around a positive 
energy density
which comes from the wall tension or the localized 
positive brane tension. 
On the other hand, $\partial_y^2U>0$ can be realized 
only if there exists a
localized artificial negative energy density \cite{RS1}. 
This physical argument suggests that the wall trajectory 
can have at most one intersecting point with
$\Omega^{(n-1)}$. 
We can prove this explicitly in the case of the 
symmetric kinetic term ($d_{1,a}=d_{1,n+1}$ for any $a$). 
For this case the BPS equation for the matter field 
can be expressed as, (see Appendix B) 
\begin{eqnarray}
\partial_y \varphi^a = \frac{1}{4}
\frac{\partial W}{\partial \varphi^a},\qquad
{\rm where}\quad\Phi^a = (1-\kappa^2m_a)(\varphi^a)^2.
\end{eqnarray}
Combining this with Eq.(\ref{BPS_warp_superpot}), 
we find 
\begin{eqnarray}
\partial_y^2U = - \frac{\kappa^2}{12}
\sum_{b=1}^{n+1}\left(
\frac{\partial W}{\partial \varphi^b}\right)^2 < 0.
\end{eqnarray}
Hence, the warp factor also has at most a single 
stationary point for our model at least in the case of 
symmetric kinetic term.

\subsection{Exact solutions}

\subsubsection{generalized $QT^*{\bf C}P^n$ 
(asymmetric kinetic term)}

First we will consider the BPS solutions for generalized 
$QT^* CP^n$ models whose kinetic terms are not symmetric 
($d_{1a}\not= d_{1b}$, $a\not=b$). 
To obtain an exact BPS solutions, we rewrite 
Eq.(\ref{BPS_Phi}) as follows:
\begin{eqnarray}
\frac{1}{\nu_{1,a}}\frac{\partial_y\Phi^a}{\Phi^a} 
= \frac{\nu_{0,a}}{\nu_{1,a}} + f(\Phi).
\end{eqnarray}
We can eliminate a complicated, but an $a$ independent 
quantity $f(\Phi)$ by subtracting ($n+1$)-th 
equation 
\begin{eqnarray}
\frac{1}{\nu_{1,a}}\frac{\partial_y\Phi^a}{\Phi^a} 
- \frac{1}{\nu_{1,n+1}}\frac{\partial_y\Phi^{n+1}}
{\Phi^{n+1}} 
= \frac{\nu_{0,a}}{\nu_{1,a}} 
-  \frac{\nu_{0,n+1}}{\nu_{1,n+1}}.
\end{eqnarray}
Then we can express $\Phi^a$ by $\Phi^{n+1}$ for 
$a=1,2,\cdots,n$:
\begin{eqnarray}
\Phi^a = \left(\Phi^{n+1}
\right)^{\frac{\nu_{1,a}}{\nu_{1,n+1}}}\ 
\exp\left[\nu_{1,a}\left(\frac{\nu_{0,a}}{\nu_{1,a}} 
-  \frac{\nu_{0,n+1}}{\nu_{1,n+1}}\right)
(y-y_a)\right],\label{Phi_n+1}
\end{eqnarray}
where $y_a$ is an integration constant. 
Plugging this into the constraint
(\ref{T_1=0}), $\Phi^{n+1}$ is implicitly obtained 
as a function of $y$:
\begin{eqnarray}
1 = \sum_{b=1}^{n+1}\frac{1}{m_b}
\left(\Phi^{n+1}\right)^{\frac{\nu_{1,b}}{\nu_{1,n+1}}}\ 
\exp\left[\nu_{1,b}\left(\frac{\nu_{0,b}}{\nu_{1,b}} 
-  \frac{\nu_{0,n+1}}{\nu_{1,n+1}}\right)
(y-y_b)\right].
\label{eq:constraint-BPS}
\end{eqnarray}
Pulling this back into Eq.(\ref{Phi_n+1}), we can also 
obtain $\Phi^a$ as a
function of $y$. 
The superpotential $W$ in 
Eq.(\ref{BPS_warp_superpot}) is given by 
\begin{eqnarray}
W 
=
 a + 
\frac{\displaystyle \sum_{b=1}^{n+1}\nu_{0,b}\Phi^b}{
\displaystyle \sum_{b=1}^{n+1}
\dfrac{1-\kappa^2m_b}{m_b}\Phi^b} , 
\label{eq:superpotential-Phi}
\end{eqnarray}
with the parameter $a$ defined in Eq.(\ref{eq:def-a}). 
The BPS equation of the function $U$ in the warp factor 
is given by Eq.(\ref{BPS_warp_superpot}) with the above 
superpotential. 
Ignoring the irrelevant integration constant of 
the warp factor, we obtain :
\begin{eqnarray}
U = - \frac{\kappa^2}{3}\ ay 
-  \frac{\kappa^2}{3}
\int dy\ \left(\frac{\displaystyle 
\sum_{b=1}^{n+1}\nu_{0,b}\Phi^b}{
\displaystyle \sum_{b=1}^{n+1}
\dfrac{1-\kappa^2m_b}{m_b}\Phi^b}\right).
\label{eq:warp-factor-final}
\end{eqnarray}
Eqs.(\ref{Phi_n+1}), (\ref{eq:constraint-BPS}), 
and (\ref{eq:warp-factor-final}) 
constitute the full set of our exact BPS solutions of 
multi-walls. 
One should note that the above derivation does not rely 
on whether there are vacuum degeneracy or not. 
Therefore the above set of solutions 
is valid, irrespective of vacua being degenerate or 
nondegenerate.  

The parameter $a$ in  Eq.(\ref{eq:warp-factor-final}) 
determines 
the behavior of the function $U$ in the  warp factor 
at infinity \cite{AFNS}. 
Since $m_1\nu_{0,1}/(1-\kappa^2m_1) \le 
m_{n+1}\nu_{0,n+1}/(1-\kappa^2m_{n+1})$ because of 
the vacuum ordering, 
we have only three possible types of asymptotic behaviors 
with respect to the AdS/CFT 
correspondence \cite{AdS/CFT}: UV-IR, IR-flat, IR-IR. 
If $0<a+m_1\nu_{0,1}/(1-\kappa^2m_1)$, we have an ultraviolet 
(UV) vacuum at $y \rightarrow -\infty$, and an infrared 
(IR) vacuum at $y\rightarrow \infty$. 
If $a+m_1\nu_{0,1}/(1-\kappa^2m_1)=0$, we have a flat 
Minkowski space at $y \rightarrow -\infty$, and an IR 
vacuum at $y\rightarrow \infty$. 
If $a+m_1\nu_{0,1}/(1-\kappa^2m_1)<0<
a+m_{n+1}\nu_{0,n+1}/(1-\kappa^2m_{n+1})$, we have an IR 
vacuum at $y \rightarrow -\infty$, and an 
IR vacuum at $y\rightarrow \infty$. 
If $a+m_{n+1}\nu_{0,n+1}/(1-\kappa^2m_{n+1})=0$, we have an IR 
vacuum at $y \rightarrow -\infty$, and  a flat 
Minkowski space at $y\rightarrow \infty$. 
If $a+m_{n+1}\nu_{0,n+1}/(1-\kappa^2m_{n+1})<0$, we have an UV 
vacuum at $y \rightarrow -\infty$, and  an IR vacuum 
at $y\rightarrow \infty$.

Our solutions contain $n$ real parameters, 
$(y_1,y_2,\cdots,y_n)$. 
They corresponds to the $n$ collective coordinates 
discovered in the BPS multi-wall solutions of 
$T^*{\bf C}P^n$ models with global SUSY \cite{GTT}. 
In the case of nondegenerate vacua, they are related 
to the locations of $n$ walls at least for large 
positive separation between them. 
If there are degenerate vacua, the parameters associated 
with these degenerate vacua actually reduces to the moduli 
parameters of the degenerate vacua. 
We shall see both these points 
more explicitly for the model with 
symmetric kinetic terms below. 
In the $T^*{\bf C}P^n$ model with global SUSY, 
there are also $n$ moduli coming from the spontaneously 
broken internal $U(1)^n_{\rm F}$ flavor symmetries 
in addition to the $n$ moduli of the wall locations 
\cite{GTT}. 
Our model also has these $U(1)^n_{\rm F}$ flavor symmetries 
which are broken spontaneously. 
This point can be seen by the freedom to choose the phase 
$\theta_a$ of our solution $\phi_1{^a}$, $a=1, \cdots, n+1$ 
in Eq.(\ref{eq:ansatz-phi1}). 
At a glance, there appear to be $n+1$
additional moduli parameters. 
In our model with SUGRA, however, we have two 
$U(1)$ directions $t_1, t_0$ which are locally gauged. 
Therefore phase rotations along these directions can be 
gauged away and unphysical. 
The phase rotation with  $\theta_a
= d_{1,a}\theta\ (a=1,2,\cdots,n+1)$ is gauged by $t_1$ 
and is absorbed into gauge field $W_\mu^1$. 
Another phase rotation with $\theta_a = d_{0,a}\theta$ is 
gauged by $t_0$ and is absorbed into gauge field $W_\mu^0$. 
Therefore, the number of the physical parameters 
associated with the internal flavor symmetries are 
$n-1$. 
Therefore the number of the massless scalar fields 
associated with the phase rotations is less by one 
than the global SUSY model. 
In the weak gravity limit, the scalar field absorbed 
by the central charge vector multiplet $W_\mu^0$ 
is recovered, since the gauge coupling $g_0$ 
should vanish and 
the associated gauge symmetry becomes global symmetry 
as $\kappa \rightarrow 0$.

Similarly to the phase rotations, there are $n$ parameters 
$y_a, a=1, \cdots, n$ corresponding to the locations of the 
walls in the global SUSY model without gravity. 
Therefore there are $n$ real parameters (collective coordinates) 
in the solution. 
In the presence of gravity, the center of mass coordinate 
is gauged away, since the gravity can be understood as a 
gauge theory for translation symmetry. 
In fact, we have recently shown in a similar SUGRA model 
in the four-dimensional spacetime 
\cite{EMS} that the scalar fluctuations 
in the Newton gauge have a zero mode which can however 
be gauged away and unphysical. 
Moreover this mode turned out to become the Nambu-Goldstone 
mode in the limit of vanishing gravitational 
coupling \cite{EMS}. 
Consequently, the physical parameters of the BPS solution 
in our model with SUGRA should be $n-1$ $U(1)$ phases and 
$n-1$ relative distances of walls, which will 
eventually form 
$n$ complex massless fields in accord with the remaining 
${\cal N}=1$ SUSY.

\subsubsection{$QT^*{\bf C}P^n$ model 
(symmetric kinetic term)}

In the case of symmetric kinetic term 
($d_{1,1}=\cdots=d_{1,n+1}$), 
it is most convenient to use the parameters 
\begin{eqnarray}
\mu_{a} \equiv \nu_{0,n+1} - \nu_{0,a}. 
\end{eqnarray}
instead of $\nu_{0, a}$. 
In this case, 
we can simplify our solution for $\Phi^a$. 
Eq.(\ref{eq:constraint-BPS}) can be inverted 
explicitly to give 
\begin{eqnarray}
\Phi^{n+1}(y) = \frac{m}{\displaystyle 
\sum_{b=1}^{n+1}{\rm e}^{-\mu_{b}(y-y_b)}} .
\label{eq:phi-n+1-sym}
\end{eqnarray}
Eq.(\ref{Phi_n+1}) can also be solved explicitly to give 
:
\begin{eqnarray}
\Phi^a(y) = \frac{m {\rm e}^{-\mu_{a}(y-y_a)}}
{\displaystyle 
\sum_{b=1}^{n+1}{\rm e}^{-\mu_{b}(y-y_b)}},\quad
\left(a=1,2,\cdots,n\right).
\label{eq:phi-a-sym}
\end{eqnarray}
The function $U$ in the warp factor 
in Eq.(\ref{eq:warp-factor-final}) 
is given more explicitly by 
\begin{eqnarray}
U = - \frac{\kappa^2}{3}
\left[
\tilde a y + \frac{m}{1-\kappa^2m}
\log\left(\sum_{b=1}^{n+1}{\rm e}^{-\mu_b(y-y_b)}\right)
\right]
\label{eq:warp-factor-sym-kin}
\end{eqnarray}
where $\tilde a\equiv a+m\nu_{0,n+1}(1-\kappa^2m)^{-1}$ 
with $a$ defined in (\ref{eq:def-a}). 
This warp factor for $n$ nondegenerate vacua 
consists of $n+1$ regions with different 
values of derivatives at least for large separations 
between $n$ walls. 
These solutions (\ref{eq:phi-a-sym}) and 
(\ref{eq:warp-factor-sym-kin}) 
are valid irrespective of vacuum degeneracy. 
However, the physical meaning of the parameters 
$y_a, \; a=1, \cdots, {n+1}$ is clearest in the case of 
nondegenerate vacua. 
At least for large positive separation, these parameters 
have an intuitive physical meaning as the 
locations of boundaries 
between vacua $a$ and $n+1$ as we shall see below. 
On the other hand, the parameter $\mu_a$ 
yields the inverse width of each wall associated 
with $y_a$. 
More precise identification will be made by taking 
$n=2, 3$ as illustrative examples below. 

We can rewrite the above 
solutions (\ref{eq:phi-n+1-sym}) and (\ref{eq:phi-a-sym}) 
for the case of degenerate vacua, 
to clarify the meaning of the parameter $y_a$ for 
degenerate vacua. 
Let us assume that the 
$\ell+1$ vacua $a, a+1, \cdots, a+\ell$ are 
degenerate and all the other vacua are nondegenerate. 
As we have seen in sect.\ref{sc:susyvac}, 
these degenerate vacua correspond 
to $\ell$ subsimplex $\sigma^{(\ell)}_1$ of 
$\Sigma^{(n)}_1$. 
If vacua at $y=\pm\infty$ are $a=I$ and $a=F$, 
all fields $\Phi^a$ with $a<I$ and $F<a$ vanish identically. 
Therefore 
we shall consider the multi-wall solution interpolating 
between the first $a=1$ and the last vacua $a=n+1$,  
without loss of generality. 
Consequently the multi-wall solution comes close to 
(or starts at, or ends at) the 
degenerate vacuum subsimplex $\sigma^{(\ell)}_1$. 
From the BPS equation, we find 
\begin{eqnarray}
\frac{\partial_y\Phi^a}{\Phi^a} 
= \frac{\partial_y\Phi^{a+1}}{\Phi^{a+1}} 
= \cdots
= \frac{\partial_y\Phi^{a+\ell}}{\Phi^{a+\ell}}.
\end{eqnarray}
This implies that the solution is embedded in an 
$(n-\ell)$-dimensional hypersurface inside 
the simplex $\Sigma^{(n)}_1$ defined as 
\begin{eqnarray}
\Phi^a = k_a\tilde \Phi^a, 
\quad 
\Phi^{a+1} = k_{a+1}\tilde\Phi^a,
\cdots, 
\Phi^{a+\ell} = k_{a+\ell}
\tilde\Phi^a, 
\end{eqnarray}
\begin{eqnarray}
\tilde \Phi^a\equiv \Phi_a+\cdots+\Phi_{a+\ell}.
\end{eqnarray}
where $k_a, \dots, k_{a+\ell}$ are arbitrary positive 
constants constrained by 
$
k_a+
\cdots 
+k_{a+\ell}=1.
$

This shows that the relative amount $k_b$ of fields 
$\Phi^b$ compared to the sum of fields associated 
to the degenerate vacua are constant. 
The remaining procedure for 
obtaining the solution is completely the same as 
above nondegenerate case. 
We find that the multi-wall solution for the $\ell+1$ 
degenerate vacua is exactly the same as the multi-wall 
solution for the case of $QT^*{\bf C}P^{n-\ell}$ 
model, with one of the field being the 
sum of these fields with degenerate vacua 
$\tilde \Phi^a$. 
\begin{eqnarray}
\tilde\Phi^a &=& 
{\rm e}^{-\mu_{a}(y-y_a)}\Phi^{n+1},
\label{dege_phi}\\
\Phi^b &=& {\rm e}^{-\mu_{b}(y-y_b)}\Phi^{n+1},
\quad (b=1, \cdots, a-1, a+\ell+1,\cdots,n+1). 
\label{nondege_phi}
\end{eqnarray}
Plugging this into the constraint 
$\displaystyle \sum_{b-1}^{n+1}\Phi^b= m$, we obtain 
\begin{eqnarray}
\Phi^{n+1} = \frac{m}{\displaystyle 
 \sum_{b=1}^{a-1}{\rm e}^{-\mu_{b}(y-y_b)}
+{\rm e}^{-\mu_{a}(y-y_a)}
+ \sum_{b=a+\ell+1}^{n+1}
{\rm e}^{-\mu_{b}(y-y_b)}}. 
\end{eqnarray}
We see that the parameters $y_a, \cdots, y_{a+\ell}$ 
in the general solutions (\ref{Phi_n+1}) and 
(\ref{eq:constraint-BPS}) are transformed into 
these relative amount $k_a$ and $y_a$ associated 
to the location of the wall. 
Clearly the field space for this degenerate vacua 
is isomorphic to the field space $\Sigma_1^{(n-\ell)}$ 
for the $QT^*{\bf C}P^{n-\ell}$ model. 
The function $U$ in the warp factor is also given by 
(\ref{eq:warp-factor-sym-kin}) with the sum of 
the fields associated with the degenerate vacua replaced by 
$\tilde \Phi^a$. 

It is now obvious how to generalize the above result to more 
involved situation with arbitrary numbers of 
groups of degenerate vacua together with nondegenerate 
vacua. 
We shall illustrate below these cases in explicit solutions 
for $n=2$ and $3$.

\subsubsection{$n=2$ case ($QT^*{\bf C}P^2$)}

\begin{figure}[h]
\centering
\hspace{-1cm}
\begin{minipage}{.45\linewidth}
\includegraphics{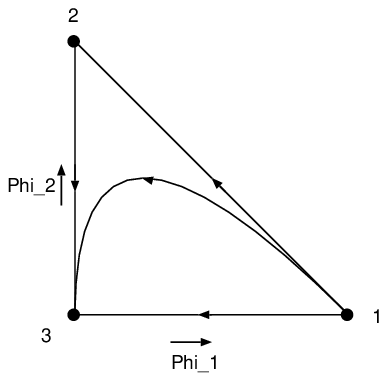}
\end{minipage}\hspace{-3cm}
\begin{minipage}{.45\linewidth}
\includegraphics[width=10cm]{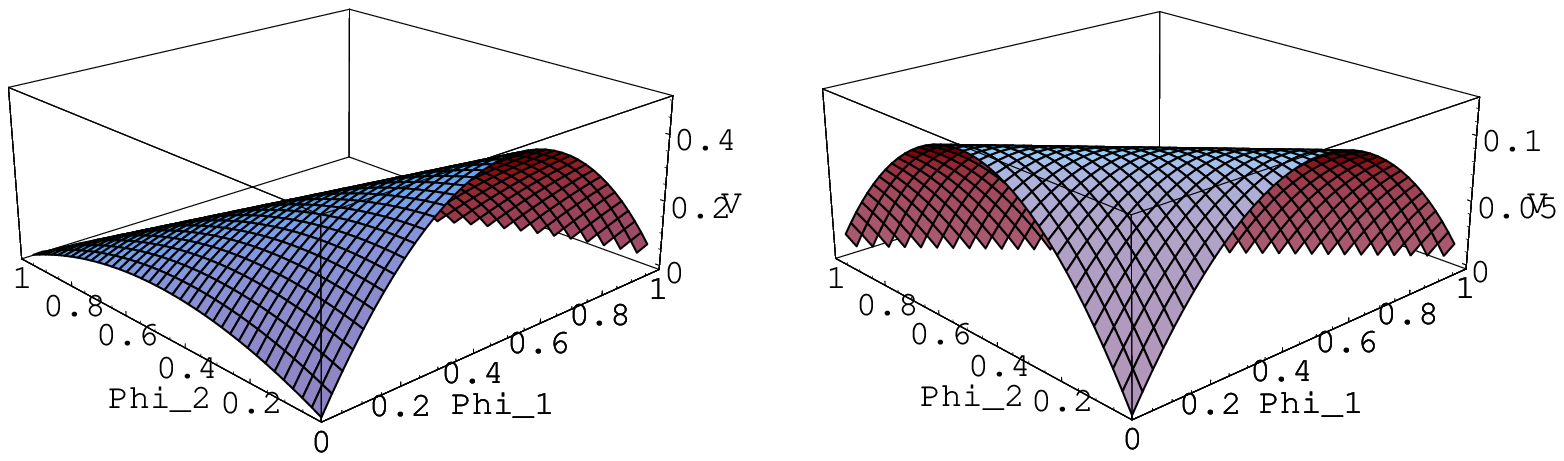}
\end{minipage}
\caption{SUSY vacua and a wall trajectory are expressed in the left
 figure. Scalar potential $V$ as a
 function of independent fields $\Phi^1, \Phi^2$ in the middle and right
 figure. The middle has $\mu_1=2\neq\mu_2=1$ (asymmetric kinetic term) and
 the right has $\mu_1=\mu_2=1$ (symmetric kinetic term). 
 In both figures $m=1, \kappa=0.1$. }
\label{fig:n2_irr}
\end{figure}

Here we deal with the $n=2$ case in some detail. 
For simplicity, we consider the case of the symmetric 
kinetic term ($d_{1,a}=d_{1,n+1}$ for any $a$) 
in the remainder of 
this section.
Let us first take nondegenerate case. 
The SUSY vacuum structure and the scalar potential are 
illustrated in 
Fig.\ref{fig:n2_irr}.

Notice that $\mu_{1}>\mu_{2}>0$ from the above 
vacuum ordering (\ref{eq:vac-ordering}). 
The general two wall solution is given by 
:
\begin{eqnarray}
\Phi^1 &=& \dfrac{m\ {\rm e}^{-\mu_{1}(y-y_1)}}{
1 + {\rm e}^{-\mu_{1}(y-y_1)} + {\rm e}^{-\mu_{2}(y-y_2)}}
=\frac{m}{1 + {\rm e}^{\mu_1(y-y_1)} 
+ {\rm e}^{(\mu_1-\mu_2)(y-y_{12})}},\\
\Phi^2 &=& \dfrac{m\ {\rm e}^{-\mu_{2}(y-y_2)}}{
1 + {\rm e}^{-\mu_{1}(y-y_1)} 
+ {\rm e}^{-\mu_{2}(y-y_2)}}
=\frac{m}{1 + {\rm e}^{\mu_2(y-y_2)} 
+ {\rm e}^{-(\mu_1-\mu_2)(y-y_{12})}},\\
\Phi^3 &=& 
\dfrac{m}{1 + {\rm e}^{-\mu_{1}(y-y_1)} 
+ {\rm e}^{-\mu_{2}(y-y_2)}}.
\label{eq:2wall-sol}
\end{eqnarray}
where we define the 
separation 
$y_{ab}$ between vacua $a$ and $b$ as 
\begin{eqnarray}
y_{ab} \equiv \dfrac{\mu_a}{\mu_a - \mu_b} y_a 
- \frac{\mu_b}{\mu_a - \mu_b}y_b,
\qquad 
y_{ab}=y_{ba}, 
\label{eq:relative-distance}
\label{eq:location-ab-wall}
\end{eqnarray}
with the understanding that $y_a \equiv y_{a, n+1}$ 
($n=2$ in our case). 
The parameter $y_{ab}$ has a physical meaning of the 
location of the boundary between the vacua $a$ and $b$. 
If these two vacua are adjacent, it in fact 
becomes the location 
of the wall separating these two vacua. 
Here, $\Phi^3$ is not an independent 
field because of the constraint $\Phi^1+\Phi^2+\Phi^3=m$.

\begin{figure}[htb]
\begin{center}
\includegraphics[width=15cm]{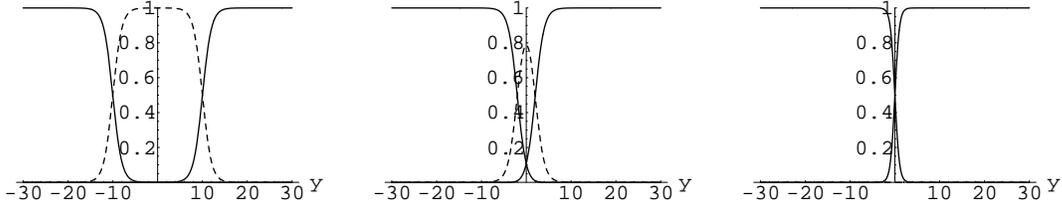}
\caption{Two-wall solutions for $n=2$ nondegenerate case : 
$y_{12}$ goes through $y_2$ from left to right. 
Parameters are taken to 
be $m=1$, and 
$(\mu_1,\mu_2)=(2,1)$, and $(y_1,y_2)=(0,10),(0,2),(0,-10)$ 
from left to right.}
\label{n2_nd}
\end{center}
\end{figure}
\begin{figure}[htb]
\begin{center}
\includegraphics[width=15cm]{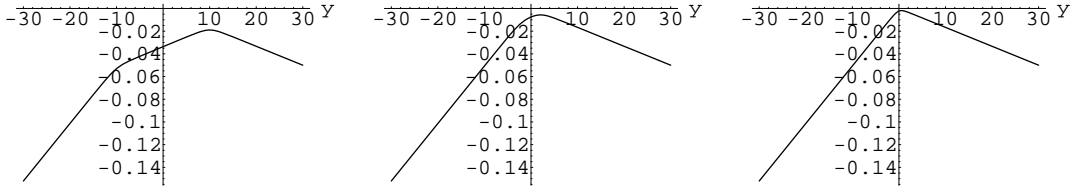}
\caption{The functions $U$ in the warp factor 
corresponding to the solutions 
in Fig.\ref{n2_nd}. 
Parameters are taken to be $\kappa=0.1$, $\tilde a=0.5$, 
 $m=1$, and 
$(\mu_1,\mu_2)=(2,1)$, and $(y_1,y_2)=(0,10),(0,2),(0,-10)$ 
from left to right. 
}
\label{fig:warp_n2}
\end{center}
\end{figure}
The solution shows that 
$\Phi^1\rightarrow m, $ and 
$\Phi^2, \Phi^3 \rightarrow 0$ 
as $y\rightarrow-\infty$,
and 
$\Phi^1, \Phi^2 \rightarrow 0$ and 
$\Phi^3 \rightarrow m$ 
as $y\rightarrow\infty$. 
If we wish to see the region of vacuum $2$ appearing 
clearly, we should arrange the parameters such that 
$y_1 < y_2$. 
Then the ordering becomes 
\begin{eqnarray}
y_{12} < y_1\equiv y_{13} < y_2\equiv y_{23}.
\end{eqnarray}
Hence, $\Phi^1$ has a wall behavior 
centered around $y=y_{12}$, and $\Phi_2$ 
has two walls at the vicinity of $y_{12}$ and $y_2$, 
($\Phi^3$ has a wall centered around $y_2$) as 
illustrated in the left-most of Fig.\ref{n2_nd}. 
The corresponding solutions of the function $U$ in the 
warp factor in Eq.(\ref{eq:warp-factor-sym-kin}) 
are shown in Fig.\ref{fig:warp_n2}. 
Three separate regions are clearly visible in the left-most 
of Fig.\ref{fig:warp_n2}. In the right-most of 
Fig.\ref{fig:warp_n2}, a sharp turn-over of $U$ 
coresponds to small (almost invisible) $\Phi^2$ 
in the right-most of Fig.\ref{n2_nd}. 

The relative distance $R$ between these two walls 
is given as 
\begin{eqnarray}
R \equiv y_2 - y_{12} = 
\frac{\mu_1}{\mu_1 - \mu_2}(y_2 - y_1).
\end{eqnarray}
When $y_2$ approaches to $y_1$, 
the two walls are compressed each other. 
If $y_2$ becomes smaller than $y_1$, 
$R$ loses the intuitive physical meaning of the 
relative distance. 
It is still a parameter of the solution which has an 
interesting nontrivial dynamics \cite{To}, \cite{To2}. 
We illustrate the behavior of multi-wall solution as 
$y_{12}$ goes through $y_{2}$ from left to right 
in Fig.\ref{n2_nd}.
The above explicit solutions (\ref{eq:2wall-sol}) satisfy 
\cite{GTT}
\begin{eqnarray}
(\Phi^1)^{\mu_2}(\Phi^2)^{-\mu_1}(\Phi^3)^{\mu_1-\mu_2}
= {\rm e}^{-\mu_2(\mu_1-\mu_2)R}.
\end{eqnarray}
Hence, the relative distance $R$ determines 
the trajectory of the solution curve in $\Sigma^{(2)}_1$, 
and the center of mass coordinate 
$y_{\rm cm}\equiv (y_{12}+y_2)/2$ determines the mapping 
from the base space $y$ to the trajectory in 
$\Sigma^{(2)}_1$.
When $R$ becomes zero or negative, 
two walls are compressed each other and the 
solution curves are close to the straight line 
(subsimplex $\sigma^{(1)}_1$) 
connecting the first and the last vacua. 
In the limit of $R\rightarrow -\infty$, it should reduce to 
the so-called {\it fundamental wall} \cite{GTT} connecting 
 the first and the last vacua directly without exciting 
 $\Phi^2$ at all.
As $R(>0)$ grows, the corresponding
solution curve goes closer to the vacuum $2$. 
If we take $R$ infinity, two far away walls 
approach 
the two fundamental walls, each of which 
corresponds to the $1 \rightarrow 2$ and 
 $2\rightarrow3$ wall, respectively. 

\begin{figure}[t]
\centering
\begin{minipage}{.35\linewidth}
\includegraphics{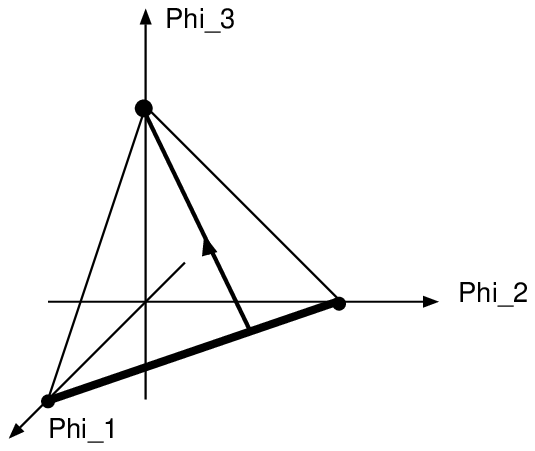}
\end{minipage}
\begin{minipage}{.35\linewidth}
\includegraphics[width=5cm]{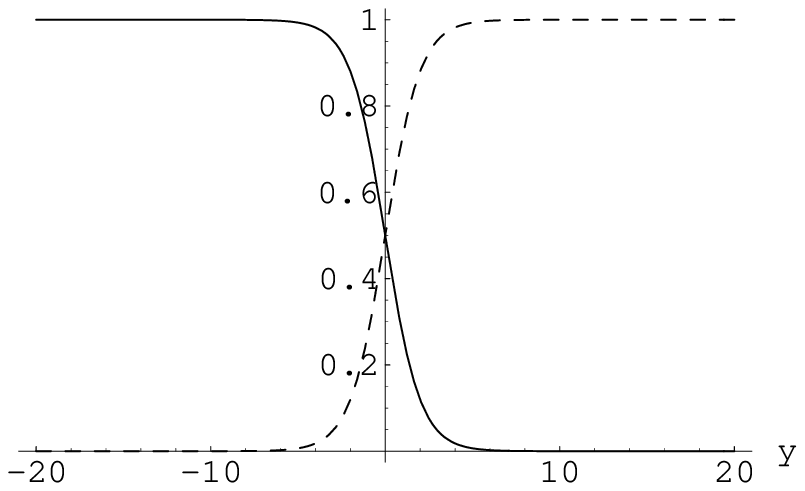}
\end{minipage}
\caption{Solution for $n=2$ degenerate case.
Parameters are taken to 
 be $m=1,\mu_1=1,y_1=0$.}
\label{n2_d}
\end{figure}

Next we turn to degenerate case with $d_{0,1} = d_{0,2}$ 
($d_{1,1}=d_{1,2}=d_{1,3}$).
In this case we have an isolated vacuum 
$(\Phi^b=m\delta^b{_3})$ and a 
degenerate vacuum which is represented by a subsimplex 
$\sigma^{(1)}_1$ with the vacua $1,2$ as its vertex. 
The BPS solutions are given by :
\begin{eqnarray}
\tilde \Phi^1(y) &\!\!\!=&\!\!\! 
\frac{m{\rm e}^{-\mu_{1}(y-y_1)}}
{1+{\rm e}^{-\mu_{1}(y-y_1)}}
=
\frac{m}
{{\rm e}^{\mu_{1}(y-y_1)}+1}
,\\
\Phi^3(y) &\!\!\!=&\!\!\! 
\frac{m}{1+{\rm e}^{-\mu_{1}(y-y_1)}}, \\
\Phi^1(y) &\!\!\!=&\!\!\! k_1 \tilde \Phi^1(y), \qquad 
\Phi^2(y) = k_2 \tilde \Phi^1(y), 
\qquad k_1+k_2=1 .
\end{eqnarray}
The solution has two parameters : $y_1$ and $k_1=1-k_2$. 
The parameter $y_1$ corresponds to the position of the
wall and $k_1$ is the relative amount of the field $\Phi^1$ 
within $\tilde \Phi^1$ associated to the degenerate vacuum. 
The field space and the solution in the case of the 
degenerate vacua is illustrated in Fig.\ref{n2_d}.

\subsubsection{$n=3$ case ($QT^*{\bf C}P^3$)}

\begin{figure}[t]
\begin{center}
\includegraphics[width=6cm]{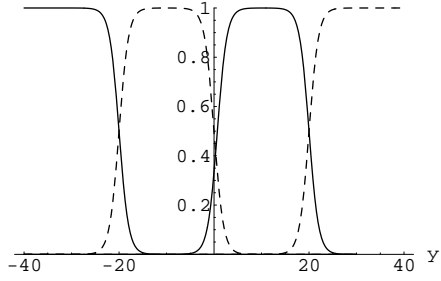}
\caption{Solution for $n=3$ nondegenerate case. 
Parameters are taken to be $m=1$, 
$(\mu_1,\mu_2,\mu_3)=(3,2,1)$ and $(y_1,y_2,y_3)=(0,10,20)$.}
\label{n3_nd}
\end{center}
\end{figure}

Here we work out the $n=3$ case in some detail. 
We shall consider the case of the symmetric kinetic term 
$d_{1,1}=d_{1,2}=d_{1,3}=d_{1,4}$.
Writing out three independent fields only, we obtain 
the three wall solution as :
\begin{eqnarray}
\Phi^1 &=& 
\frac{m}{1 + {\rm e}^{\mu_1(y-y_1)} 
+ {\rm e}^{(\mu_1-\mu_2)(y-y_{12})}
+ {\rm e}^{(\mu_1-\mu_3)(y-y_{13})}},\\
\Phi^2 &=& \frac{m}{1 + {\rm e}^{\mu_2(y-y_2)} 
+ {\rm e}^{(\mu_2-\mu_3)(y-y_{23})}
+ {\rm e}^{-(\mu_1-\mu_2)(y-y_{12})}},\\
\Phi^3 &=& \frac{m}{1 + {\rm e}^{\mu_3(y-y_3)} 
+ {\rm e}^{-(\mu_1-\mu_3)(y-y_{13})}
+ {\rm e}^{-(\mu_2-\mu_3)(y-y_{23})}},
\label{eq:3wall-sol}
\end{eqnarray}
where we used the variable $y_{ab}$ defined in 
Eq.(\ref{eq:relative-distance}). 
Notice that 
$y_{ab}$ 
are not fully independent, and satisfy the identity 
such as\footnote{
Similar identities are valid for any string of 
relative distances $y_{ab}, y_{bc}, \cdots, y_{da}$: 
$(\mu_a-\mu_b)y_{ab} + (\mu_b-\mu_c)y_{bc} 
+ \cdots + (\mu_d-\mu_a)y_{da} = 0$. 
}
 : 
\begin{eqnarray}
(\mu_1-\mu_2)y_{12} + (\mu_2-\mu_3)y_{23} 
+ (\mu_3-\mu_1)y_{31} = 0.
\end{eqnarray}
If we wish to have the region of vacuum $2$ and $3$ 
clearly visible, we should arrange the parameters 
$y_{12}<y_{23}<y_3$. 
Then we automatically obtain the ordering 
\begin{eqnarray}
y_{12} < y_{13} < y_{23} < y_2(=y_{24}) < y_3(=y_{34}).
\end{eqnarray}
Provided these relative distances are large and positive, 
$y_{12}$, $y_{23}$ and $y_3$ have an intuitive physical 
meaning of the location of 
the left (vacuum$1\rightarrow$ vacuum$ 2$) wall, the middle 
(vacuum$2\rightarrow$ vacuum$ 3$) wall, and the right 
(vacuum$3\rightarrow$ vacuum$ 4$) 
wall. 
Then the relative distances between two adjacent 
walls can be 
defined as 
\begin{eqnarray}
R_{12} = y_{23} - y_{12},\quad
R_{23} = y_3 - y_{23}.
\end{eqnarray}
A typical three wall solution for nondegenerate 
case is shown in Fig.\ref{n3_nd}.

Next we turn to the degenerate case. 
Compared to the $n=2$ case, we have more varieties 
of degenerate cases. 
Since there are four possible vacua, 
there can be degeneracies among two, three, and four vacua. 
The case of four degenerate vacua does not give wall 
solution. 
The case of three degenerate vacua can occur either 
for the first three vacua $(1, 2, 3)$ or the last three 
vacua $(2, 3, 4)$. 
The case of two degenerate vacua can occur either 
between the first two vacua $(1,2)$, or the last two 
vacua $(3,4)$, or the middle two vacua $(2,3)$. 
Consequently we can have two groups of degenerate vacua 
$(1,2)$ and $(3,4)$. 
We shall illustrate these typical solutions in 
the following. 

\begin{figure}[htb]
\begin{center}
\hspace*{-1cm}
\includegraphics{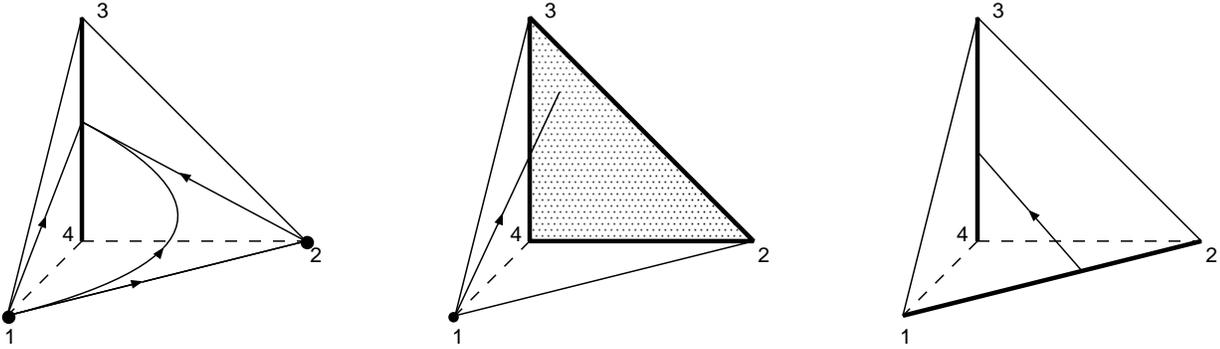}
\caption{Solution for $n=3$ degenerate case}
\label{n3_d}
\end{center}
\end{figure}

If the last two vacua $(3,4)$ are degenerate 
and the rest are nondegenerate, we obtain 
\begin{eqnarray}
\Phi^1 &=& 
\frac{m}{1 + {\rm e}^{\mu_1(y-y_1)} 
+ {\rm e}^{(\mu_1-\mu_2)(y-y_{12})}},\\
\Phi^2 &=& 
\frac{m}{1 + {\rm e}^{\mu_2(y-y_2)} 
+ {\rm e}^{-(\mu_1-\mu_2)(y-y_{12})}},
\\
\tilde \Phi^3 &\equiv& \Phi^3+\Phi^4
=\dfrac{m}{1 + {\rm e}^{-\mu_{1}(y-y_1)} 
+ {\rm e}^{-\mu_{2}(y-y_2)}}, 
\\
\Phi^3&=&k_3\tilde \Phi^3, \qquad
\Phi^4=k_4\tilde \Phi^3, \qquad k_3+k_4=1
.
\label{eq:2wall-sol-dege}
\end{eqnarray}
This is the two-wall solution 
which is the same as 
that in Eq.(\ref{eq:2wall-sol}) for the 
$QT^*{\bf C}P^2$ model, except 
that the field $\Phi^3$ in Eq.(\ref{eq:2wall-sol}) 
is now replaced by $\tilde \Phi^3$, and 
the ratio $k_3/k_4=k_3/(1-k_3)$ 
of the two fields $\Phi^3/\Phi^4$ is another 
parameter of the solution. 
This case is illustrated in Fig.\ref{n3_d}(a).
The case of first two vacua being degenerate is very similar. 
If the middle two vacua are degenerate, we can use the 
same solution for $QT^*{\bf C}P^2$ in Eq.(\ref{eq:2wall-sol}) 
by replacing the middle field $\Phi^2$ with 
the sum of the fields $\tilde \Phi^2 \equiv \Phi^2+\Phi^3$ 
associated to the degenerate vacua and the last field by 
$\Phi^4$. 
The ratio $k_2/k_3=k_2/(1-k_2)$ 
of these two fields $\Phi^2/\Phi^3$ is another 
parameter of the solution. 
The wall trajectory in field space 
is similar to Fig.\ref{n3_d}(a). 

If the last three vacua ($2, 3, 4$) are degenerate, 
we obtain a single wall solution 
\begin{eqnarray}
\Phi^1 &=& \dfrac{m\ }{
1 + {\rm e}^{\mu_{1}(y-y_1)}}
,\\
\tilde \Phi^2  &\equiv& \Phi^2+\Phi^3+\Phi^4
= \dfrac{m\ }{
1 + {\rm e}^{-\mu_{1}(y-y_1)} 
}
,
\\
\Phi^2&=&k_2\tilde \Phi^2, 
\quad 
\Phi^3=k_3\tilde \Phi^2, \quad
\Phi^4=k_4\tilde \Phi^2, \quad k_2+k_3+k_4=1
.
\label{eq:1wall-sol-dege}
\end{eqnarray}
This solution is 
the same as that in 
Eq.(\ref{eq:2wall-sol}) for the $QT^*{\bf C}P^2$ model 
with $\tilde \Phi^2$ replacing $\Phi^2$. 
The relative amount $k_a$ of $\Phi^a$ in $\tilde \Phi^2$ 
are another parameter of the solution. 
This case is illustrated in Fig.\ref{n3_d}(b).
The case of last three vacua being degenerate is 
very similar. 

Another interesting case is the two groups of 
degenerate vacua $(1,2)$ and $(3, 4)$. 
We obtain a single wall solution with 
$\tilde \Phi^1$ replacing $\Phi^1$ and 
$\tilde \Phi^3$ replacing $\tilde \Phi^2$ in the 
above solution (\ref{eq:1wall-sol-dege}). 
The relative amount $k_a$ of $\Phi^a$ in $\tilde \Phi^2$ 
are also the parameters of the solution 
\begin{equation}
\Phi^1=k_1\tilde \Phi^1, 
\quad 
\Phi^2=k_2\tilde \Phi^1, \quad k_1+k_2=1, 
\qquad 
\Phi^3=k_3\tilde \Phi^3, \quad 
\Phi^4=k_4\tilde \Phi^3, \quad 
k_3+k_4=1
. 
\end{equation}
This case is illustrated in Fig.\ref{n3_d}(c).

\section{
Discussion}
\label{sc:discuss}

Let us first discuss weak gravity limit. 
In taking the limit, we demand that 
meaningful models are obtained with the global SUSY. 
As we noted, we should fix $g_0M^0$ and $\alpha_0/\kappa^2$ 
and $\alpha_1/\kappa^2$ when taking the limit of 
vanishing gravitational coupling $\kappa\rightarrow 0$ 
\cite{AFNS}. 
By taking this limit with the symmetric kinetic term, 
we can recover the global SUSY model with the target 
space $T^*{\bf C}P^n$ which was studied before 
in lower dimensions \cite{GTT}. 
In this limit, BPS equations for scalar fields 
in the hypermultiplets reduce to those already studied 
before \cite{GTT}. 

Therefore the model and the solution we discuss in this paper 
are consistent gravitational deformation 
of the massive $T^*{\bf C}P^n$ nonlinear sigma model 
in five dimensions and associated BPS wall solutions. 
It is very interesting that 
BPS solution for the hypermultiplet $\phi$ 
in the global SUSY model coincides with 
that in the corresponding SUGRA. 
This mysterious coincidence was noted in the case of 
massive Eguchi-Hanson ($T^*{\bf C}P^1$) nonlinear 
sigma model \cite{AFNS}. 
This situation 
has also appeared in the analytic 
solution in a four-dimensional ${\cal N}=1$ SUGRA
model \cite{EMSS}.  
It is tempting to speculate that this property 
might be related to the exact solvability of 
our model in SUGRA.

For a long time, it has been a difficult problem to 
find a consistent gravitational 
deformation from a hyper-K\"ahler manifold 
to a quaternionic K\"ahler manifold 
with gravitationally corrected potential terms 
necessary for wall solutions. 
This goal has been achieved here by using an off-shell 
formulation of SUGRA and 
the {\it massive} quaternionic K\"ahler quotient 
method \cite{AFNS}. 
There has been an extensive studies of domain walls 
in SUGRA theories, 
using the on-shell formulation such as 
in Ref.\cite{Cere-Dall}.
In principle, it may be possible to obtain BPS solutions 
using the on-shell formulation, since auxiliary fields 
are eliminated when we solve the BPS equations. 
However, off-shell formulation of SUGRA provides 
a more powerful tool to obtain SUGRA domain walls 
as gravitational deformations of those in global SUSY 
models. 
If we eliminate constraints before coupling to gravity, 
target manifold of nonlinear sigma models are fixed 
already. 
Then it is very difficult in general to find out necessary 
gravitational corrections to the target manifold 
in order to extend hyper-K\"ahler nonlinear sigma models 
with global eight SUSY to quaternionic K\"ahler nonlinear 
sigma models coupled to SUGRA.
On the other hand, many hyper-K\"ahler sigma models 
can be obtained as quotients of linear sigma models 
by using vector multiplets as Lagrange multipliers 
\cite{LR}, \cite{CF}, \cite{ANNS}, \cite{ANNS2}. 
We can first couple these system of gauged linear sigma 
models to SUGRA before 
eliminating the Lagrange multiplier multiplets. 
When we eliminate the Lagrange multipliers 
after coupling to 
gravity in the off-shell formulation, 
we obtain quaternionic K\"ahler nonlinear sigma models 
coupled to SUGRA automatically. 
In this framework, we can take a weak gravity limit 
of these models straightforwardly. 
Therefore the off-shell formulation of SUGRA
is very useful 
to obtain quaternionic nonlinear sigma models as continuous 
gravitational deformations of hyper-K\"ahler nonlinear 
sigma models of the global SUSY. 

By construction, our models should have quaternionic 
K\"ahler target manifold as far as local properties are 
concerned. 
However, we have not yet studied possible global 
obstructions which might exist in these target manifold. 
We still have to work out the coordinates that parametrize 
the manifold globally. 
This is a problem to be studied in future. 

At least in the case of $T^*{\bf C}P^1$ model, 
it has been noted that 
our quaternionic manifold has a 
conical singularity at $r=0$ in $r, \Psi$ plane 
\cite{AFNS}, 
except for discrete values 
of gravitational coupling 
$\kappa^2\Lambda^3=(k-1)/k, \; k=2,3,\dots$ 
where it can be identified with a 
removable bolt singularity \cite{IvanovValent}. 
Here one uses ${\bf Z}_k$ division 
instead of ${\bf Z}_2$ of the original 
Eguchi-Hanson metric. 
However, we find it more desirable to have a smooth 
limit of vanishing gravitational coupling. 
In this respect, it is strange to use different divisions 
for a model with global SUSY and its gravitationally 
deformed version. 
To achieve a smooth limit of vanishing gravitational 
coupling, our strategy to deal with such singularities 
is to restore the kinetic term for the Lagrange multiplier 
multiplet $W_\mu^1$ \cite{AFNS}. 
We believe that we can 
still achieve a continuous 
gravitational deformation avoiding the singularity, 
if we simply restore 
the kinetic term for the vector multiplet 
containing $W_\mu^1$. 
The elimination of the kinetic term correspond to the 
infinite gauge coupling $g_1\rightarrow \infty$. 
Let us take a gauged linear sigma model consisting of 
hypermultiplets interacting with vector multiplets and 
couple it to SUGRA by the tensor calculus 
\cite{Fujita-Ohashi}. 
This model is a consistent interacting 
SUGRA system with eight local SUSY. 
For finite but large values of gauge coupling $g_1$, 
the model effectively reduces 
to our quaternionic nonlinear sigma model except near the 
conical singularity where we can no longer neglect the 
vector multiplet. 
Only in the neighborhood of the singularity, the manifold 
loses its simple geometrical meaning of quaternionic 
manifold consisting solely of 
hypermultiplets. 
In this gauged linear sigma 
model, we can freely take 
the limit $\kappa \rightarrow 0$ to obtain 
the $QT^*{\bf C}P^n$ manifold. 
Therefore we believe that this gauged linear 
sigma model coupled with SUGRA
is most suitable to obtain the gravitational deformation of 
hyper-K\"ahler manifolds. 
On the other hand, our BPS multi-wall solutions 
should still be valid solutions of the 
gauged linear sigma model coupled with SUGRA.
This is because 
our constraints arising from the 
elimination of the vector multiplet without kinetic 
term preserve all SUSY, and hence they 
solve the additional BPS condition 
for the vector multiplet 
trivially. 
Therefore we anticipate that our solution continues 
to be a BPS wall solution 
for the gauged linear sigma model with a finite large 
gauge coupling $g_1$ coupled with SUGRA.
The only modification should be that 
near the conical singularity, the 
target space manifold loses a simple geometrical meaning as 
a genuine quaternionic manifold consisting solely of 
hypermultiplets. 
Instead the vector multiplet cannot be neglected 
when we examine the geometry of the target manifold 
near the ``resolved'' conical singularity \cite{Witten}. 
A full analysis of the gauged linear 
sigma model 
with a finite gauge coupling $g_1$ (with the kinetic term) 
coupled to SUGRA is under investigation. 

Since we have vector multiplets $W_\mu^1$ and $W_\mu^0$, 
we should take into account the possible Higgs mechanism 
due to these gauge fields. 
Even in the limit of no kinetic term for $W_\mu^1$ 
as in our case here, the gauge field $W_\mu^0$ 
associated with the central charge extension 
should absorb one of the Nambu-Goldstone bosons 
associated with the spontaneously broken $U(1)_F^n$ 
global flavor symmetry. 
This is a new feature of our model with SUGRA
compared to the $T^*{\bf C}P^n$ model with global SUSY. 
We note, however, this Higgs mechanism should be 
absent in the limit of vanishing gravitational coupling. 
This is ensured by the fact that 
the gauge coupling $g_0$ vanishes in the limit 
of the vanishing gravitational coupling 
\begin{equation}
(M^0)^2={3 \over 2\kappa^2}, 
\quad 
g_0 \sim {1 \over M^0} \sim {\kappa} \rightarrow 0, 
\qquad 
\kappa \rightarrow 0. 
\end{equation}
We shall study also these aspects associated with 
the vector multiplets in separate publications.


\renewcommand{\thesubsection}{Acknowledgments}
\subsection{}

One of the authors (N.S.) is indebted to useful discussion 
with Keisuke Ohashi, Sergei Ketov, 
and 
Taichiro Kugo.
This work is supported in part by Grant-in-Aid for Scientific 
Research from the Ministry of Education, Culture, Sports, 
Science and 
Technology,Japan, 
 No.13640269. 
One of the authors (M.E.) gratefully acknowledges 
support from the Iwanami Fujukai Foundation.

\renewcommand{\thesubsection}{\thesection.\arabic{subsection}}

\appendix

\section{Isometry of Our Action} 
\label{isometry}

Since the generator $t_1$ gives the constraint to 
obtain the curved 
target manifold, $t_1$ determines the geometry of the 
target manifold, in particular its isometry. 
Therefore we can examine the isometry of our model 
(\ref{SUGRA1}) 
once we fix the matrix $t_1$. 
On the other hand, another $U(1)$ generator $t_0$ 
associated to the central charge multiplet gives a mass term 
which eventually specifies the potential term. 
We should choose the generator $t_0$ among the 
generators $t$ of the isometry. 

Now we want to construct $U(1) \times U(1)$ gauged action, 
$t_1$ and $t_0$ must commute each other. 
We shall choose $t_1$ as 
\begin{eqnarray}
\left(t_1\right)^\alpha{_\beta}
= \left(
\begin{array}{ccc}
i\alpha_1\sigma_3 &&\\
&i\bd{D}_1 &\\
&&-i\bd{D}_1
\end{array}
\right), 
\end{eqnarray}
where $\bd{D}_1= diag(d_{1,1},d_{1,2},\cdots,d_{1,n+1})$. 
Those generators that commute with $t_1$ form the 
isometry of the target manifold. 
We have to choose $t_0$ among generators of 
isometry. 
If we choose the other generator $t_0$ as diagonal, 
we obtain the $U(1)\times U(1)$ gauging for 
arbitrary $\alpha_1$ and $d_{1,a}$s. 
For generic values of $\alpha_1$ and $d_{1,a}$s, 
these are the only matrix commuting with the $t_1$ 
generators.  
In that case, isometry of the manifold is $U(1)^n$. 
For specific values of $\alpha_1$ and $d_{1,a}$s, 
however, $t_0$ need not be diagonal, but can have 
off-diagonal elements. 
Namely, the isometry of the target manifold is enhanced. 

Because of the gauge fixing of dilatation and 
invariance of quadratic forms, 
generators should satisfy Eq.(\ref{eq:constr-t}). 
These matrices are the generators of 
the group $USP(2,2(n+1))$, and are given explicitly as 
in Eq.(\ref{eq:USpgenerator}) 
Therefore we should choose the generators $t_0$ 
among $USP(2,2(n+1))$ generators 
\begin{eqnarray}
\left(t_0\right)^\alpha{_\beta}
 = \left(
\begin{array}{ccc}
\bd{A} & \bd{B}^\dagger & - \varepsilon\bd{B}^{\rm T}\\
\bd{B} & \tilde{\bd{D}} & \bd{C}\\
\bd{B}^*\varepsilon & \bd{C}^* & - \tilde{\bd{D}}^{\rm T}
\end{array}
\right), 
\end{eqnarray}
where $\bd{A}$ is a $2\times2$ matrix which satisfies 
$\bd{A}^\dagger =
-\bd{A}$ and $\bd{A}=\varepsilon\bd{A}^{\rm T}\varepsilon$, 
$\bd{B}$ is
a $(n+1)\times2$ matrix, $\bd{C}$ is a 
$(n+1)\times(n+1)$ matrix which
satisfies $\bd{C}^{\rm T} = -\bd{C}$, and 
$\tilde{\bd{D}}$ is
a $(n+1)\times(n+1)$ matrix which satisfies 
$\tilde{\bd{D}}^\dagger = -\tilde{\bd{D}}$. 

The condition $[ t_0,t_1 ]=0$ can be rewritten as follows: 
\begin{eqnarray} 
&\alpha_1 [\sigma_3 ,\bd{A}]=0,& \label{isoA} 
\\ 
&
\bd{D}_1 \bd{B}-\alpha_1 \bd{B} \sigma_3 =0,& 
\label{isoB} \\ 
&\{ \tilde{\bd{D}} ,\bd{C}\}=0,& \label{isoC} \\ 
&[\bd{D}_1 , \tilde{\bd{D}}]=0.& \label{isoD} 
\end{eqnarray} 

We shall concentrate on $n=1$ case here. 
For $n=1$, these relations allow 
non-diagonal $t_0$ depending on the values of 
parameters $\alpha_1, d_{1, a}$. 
The equation (\ref{isoA}) is satisfied by 
$\alpha_1=0$. 
The equation (\ref{isoB}) is satisfied by 
$\alpha_1=\pm d_{1,1}, \ \pm d_{1,2} $. 
The equation (\ref{isoC}) is satisfied by  
$d_{1,1}=0$, $d_{1,2}=0$, or 
$d_{1,1}+d_{1,2}=0$. 
The equation (\ref{isoD}) is satisfied by  
$d_{1,1}=d_{1,2}$. 
When $\alpha_1 \neq 0$, we can classify 
various cases of symmetry into the seven types 
as listed on Table \ref{table:1}. 
\begin{table}[htb]
\caption{The isometry of our model ($n=1$)}
\label{table:1}
\begin{center}
\begin{tabular}{c|c|c} \hline \hline
     & conditions   & isometry  \\ \hline \hline 
i)   & $d_{1,1}=d_{1,2}=0$ & $USp(4)$ \\ \hline 
ii)  & $\alpha_1 =\pm d_{1,1},\ d_{1,1}
=\pm d_{1,2}$ & $SU(1,2)$  \\ \hline 
iii) & $\alpha_1 =\pm d_{1,1},\ d_{1,2}=0$ 
& $SU(1,1)\times SU(2)$  \\ 
     & $\alpha_1 =\pm d_{1,2},\ d_{1,1}=0$ 
     &           \\ \hline 
iv)  & $d_{1,1}=\pm d_{1,2}\neq 0$ & $U(2)$ \\ \hline 
v)   & $d_{1,1}=0$ or $d_{1,2}=0$ & $U(2)$ \\ \hline 
vi)  & $\alpha_1 =\pm d_{1,1},\ d_{1,2}\neq 0,\ 
\pm d_{1,1}$ & $SU(1,1)\times U(1)$  \\ 
     & $\alpha_1 =\pm d_{1,2},\ d_{1,1}\neq 0,\ 
     \pm d_{1,2}$ &   \\ \hline 
vii) & otherwise & $U(1)\times U(1)$ \\ \hline 
\end{tabular}
\end{center}
\end{table} 

Among these possibilities, however, 
the case i), iii), v) are not 
desirable by the following reasons. 
When at least one of the 
$d_{1,a}$ vanishes, 
the direction $a$ is not gauged. 
Then we do not obtain a curved target manifold due to the 
constraint along this direction. 
When $\alpha_1=0$, 
the mass scale of the target manifold goes to infinity, 
and we again obtain no additional constraint through 
the vector multiplet without kinetic term. 
These are not what we want. 
In case ii), $t_1$ itself vanishes. 
Then 
the target metric becomes flat in the global limit, 
or global limit do not exist \cite{Beh-Dall}. 
Furthermore, this case is not suitable for 
constructing the wall solutions 
for our purpose 
because of divergence of the vacuum energy 
in the global limit. 

Therefore only the iv), vi), vii) cases are suitable 
to obtain quaternionic extensions 
of the hyper-K\"ahler manifolds and to construct 
the wall solutions. 
The cases iv), vi) have enhanced isometry, whereas 
the case vii) gives the smallest isometry $U(1)\times U(1)$.


\section{Superpotential}

The purpose of this appendix is to simplify our
model by introducing the ``superpotential''.
Let us first attempt to 
simplify the scalar potential (\ref{scalar_pot}) 
and the BPS equations (\ref{BPS_Phi}). 
For that purpose we introduce the following real 
vectors using the real fields 
$\varphi^a \equiv \sqrt{{\Phi^a}/{(1-\kappa^2m_a)}}
=|\phi_1^a|/\sqrt{1-\kappa^2m_a}$ 
:
\begin{eqnarray}
\bd v = \left(\nu_{0,1}\varphi^1,\cdots, 
\nu_{0,n+1}\varphi^{n+1}\right)^{\rm T},\quad
\bd w = \left(\dfrac{1-\kappa^2m_1}{m_1}\ \varphi^1,\cdots,
\dfrac{1-\kappa^2m_{n+1}}{m_{n+1}}\ 
\varphi^{n+1}\right)^{\rm T},
\end{eqnarray}
 It is also useful to introduce a matrix and vectors :
\begin{eqnarray}
\bd M^{\frac{1}{2}} \equiv diag
\left(\sqrt{m_1},\sqrt{m_2},\cdots,\sqrt{m_{n+1}}\right),
\qquad 
\bd v' = \bd M^{\frac{1}{2}} \bd v,\quad
\bd w' = \bd M^{\frac{1}{2}} \bd w.
\end{eqnarray}
Then we find the following relations :
\begin{eqnarray}
(\bd v,\bd v) &=& \sum_{b=1}^{n+1} 
\nu_{0,b}^2\ (\varphi^b)^2 
= \sum_{b=1}^{n+1} \frac{1}{1-\kappa^2m_b}\ 
\nu_{0,b}^2\ \Phi^b,\\
(\bd v,\bd w) &=& \sum_{b=1}^{n+1} 
\frac{1-\kappa^2m_b}{m_b}\ \nu_{0,b}\ (\varphi^b)^2 
= \sum_{b=1}^{n+1} \nu_{0,b}\ \frac{\Phi^b}{m_b},\\
(\bd w,\bd w) &=& \sum_{b=1}^{n+1} 
\left(\frac{1-\kappa^2m_b}{m_b}\right)^2(\varphi^b)^2 
= \sum_{b=1}^{n+1} 
\frac{1-\kappa^2m_b}{m_b}\ \frac{\Phi^b}{m_b}.
\end{eqnarray}
\begin{eqnarray}
(\bd v',\bd v') &=& \sum_{b=1}^{n+1} m_b\ 
\nu_{0,b}^2\ (\varphi^b)^2 
= \sum_{b=1}^{n+1} \frac{m_b}{1-\kappa^2m_b}\ 
\nu_{0,b}^2\ \Phi^b,\\
(\bd v',\bd w') &=& \sum_{b=1}^{n+1} 
(1-\kappa^2m_b)\ \nu_{0,b}\ (\varphi^b)^2 
= \sum_{b=1}^{n+1} \nu_{0,b}\ \Phi^b,\\
(\bd w',\bd w') &=& \sum_{b=1}^{n+1} m_b\ 
\left(\frac{1-\kappa^2m_b}{m_b}\right)^2(\varphi^b)^2 
= \sum_{b=1}^{n+1} \frac{1-\kappa^2m_b}{m_b}\ \Phi^b.
\end{eqnarray}
Using these relations, we can express the scalar potential
$V$ (\ref{scalar_pot}) as follows:
\begin{eqnarray}
V = \frac{1}{2(\bd w', \bd w')}
\left[(\bd v, \bd v)\sin^2\theta 
- \kappa^2 (\bd v',\bd v')\sin^2\theta'\right]
- \frac{2\kappa^2}{3}\left( a + 
\frac{(\bd v',\bd w')}{(\bd w',\bd w')}\right)^2,
\label{pot_simple}
\end{eqnarray}
where we define
\begin{eqnarray}
\cos\theta \equiv 
\frac{(\bd v,\bd w)}{|\bd v||\bd w|},\quad
\cos\theta' \equiv 
\frac{(\bd v',\bd w')}{|\bd v'||\bd w'|}.
\end{eqnarray}

The BPS equations can also be expressed as follows:
\begin{eqnarray}
\partial_y\varphi^a 
&=& \frac{1}{2}\left(\nu_{0,a} - \frac{(\bd v,\bd w)}{
(\bd w,\bd w)}\frac{1-\kappa^2m_a}{m_a}\right)\varphi^a,\\
\partial_y U &=& - \frac{\kappa^2}{3}
\left( a + \frac{(\bd v',\bd w')}{(\bd w',\bd w')}\right).
\end{eqnarray}
Let us introduce two real functions $W$ and $\tilde W$:
\begin{eqnarray}
W(\varphi^a) \equiv a + 
\frac{(\bd v',\bd w')}{(\bd w',\bd w')},\quad
\tilde W(\varphi^a) \equiv a 
+ \frac{(\bd v,\bd w)}{(\bd w, \bd w)}.
\end{eqnarray}
The above scalar potential can be expressed 
in terms of these functions as :
\begin{eqnarray}
V = \frac{1}{8}\sum_{b=1}^{n+1}\left((\bd w,\bd w)
\frac{m_a}{1-\kappa^2m_a}\right)
\frac{\partial \tilde W}{\partial \varphi^b}
\frac{\partial W}{\partial \varphi^b}
- \frac{2\kappa^2}{3}W^2.\label{pot_suppot}
\end{eqnarray}
Notice that this expression is valid for general 
asymmetric kinetic term. 
The BPS equations can also be expressed as follows :
\begin{eqnarray}
\partial_y\varphi^a = \frac{1}{4}\left((\bd w,\bd w)
\frac{m_a}{1-\kappa^2m_a}\right)
\frac{\partial \tilde W}{\partial\varphi^a},\quad
\partial_y U = - \frac{\kappa^2}{3}W.
\end{eqnarray}
The nontrivial target space is realized through 
the constraint 
\begin{eqnarray}
1 = \sum_{b=1}^{n+1}
\frac{1-\kappa^2m_b}{m_b}(\varphi^b)^2.
\end{eqnarray}

These expressions reduce to familiar forms when all the
$m_a$ are equal $(=m)$ to each other as in the case of 
symmetric kinetic terms. 
Since $W = \tilde W$ and $(\bd w,\bd w) =
\dfrac{1-\kappa^2 m}{m}$ in this case, we find
\begin{eqnarray}
V = \frac{1}{8}\sum_{b=1}^{n+1} 
\left(\frac{\partial W}{\partial \varphi^b}\right)^2
- \frac{2\kappa^2}{3}W^2,
\label{eq:potential-SUGRA}
\\
\partial_y\varphi^a = 
\frac{1}{4}\frac{\partial W}{\partial \varphi^a},\quad
\partial_yU = - \frac{\kappa^2}{3}W.
\end{eqnarray}
If we are interested in the section of field space 
with $\phi_2^a=0$ in 
Eq.(\ref{eq:ansatz-phi1}) and 
the constant phase Ansatz for $\phi_1^a$ in 
Eq.(\ref{eq:ansatz-phi1}), we find the bosonic part 
of the action to be rewritten with these fields $\varphi^a$ 
as 
\begin{eqnarray}
S = \int d^{5}x \ e\left[
- \frac{1}{2\kappa_5^2}R - 
2\sum_i\left(\partial_\mu\varphi^a\right)^2 - V(\varphi^a)
\right]. 
\label{eq:eff-action-SUGRA}
\end{eqnarray}

It has been known that the scalar potential 
of many SUGRA theories can be expressed by a real
function $\hat W(\varphi^i)$, the so-called ``superpotential'', as 
\cite{BS, Skenderis:1999mm, DFGK} 
\begin{eqnarray}
V(\varphi^a) = {(D-2)^2 \over 2}\sum_i
\left(\frac{\partial \hat W(\varphi^a)}{\partial\varphi^a}\right)^2
- 2(D-1)(D-2)\kappa_D^2\hat W^2(\varphi^a),
\end{eqnarray}
where the scalar fields are normalized such that 
the $D$-dimensional Einstein gravity and a number of 
scalar fields $\varphi^a$ with a scalar potential $V$ 
is given by : 
\begin{eqnarray}
S = \int d^{D-1}xdy\ e\left[
- \frac{1}{2\kappa_D^2}R - 
2\sum_i\left(\partial_\mu\varphi^a\right)^2 - V(\varphi^a)
\right],
\end{eqnarray}
where $\kappa_D^2 = \dfrac{1}{M_D^{D-2}}$ 
and $e$ is the determinant of
the vielbein. 
This form of the scalar potential ensures the existence 
of the stable AdS 
vacua with or without SUSY, if the superpotential 
has at least a stationary point \cite{Boucher, PKT}.
By a rescaling $W=6\hat W$, 
our results 
(\ref{eq:potential-SUGRA})--(\ref{eq:eff-action-SUGRA}) 
are consistent with 
this ``effective SUGRA'' 
formalism in $D=5$ dimensions.

\end{document}